\documentclass[pra,twocolumn,prl,groupedaddress]{revtex4-1}

\usepackage{graphicx}
\usepackage{dcolumn}
\usepackage{bm}
\usepackage{epsfig}
\usepackage{amsmath}
\usepackage{amssymb}
\usepackage{color}
\usepackage{bbm}
\usepackage{braket}

\usepackage[colorlinks=true,citecolor=blue,linkcolor=blue,urlcolor=blue]{hyperref}

\begin{document}

\title{Breakdown of gauge invariance in ultrastrong-coupling cavity QED}

\author{Daniele De Bernardis$^1$, Philipp Pilar$^1$, Tuomas Jaako$^1$, Simone De Liberato$^2$, and Peter Rabl$^1$}
\affiliation{$^1$Vienna Center for Quantum Science and Technology, Atominstitut, TU Wien, 1040 Vienna, Austria}
\affiliation{$^2$School of Physics and Astronomy, University of Southampton, Southampton, SO17 1BJ, United Kingdom}

\date{\today}

\begin{abstract} 
We revisit the derivation of Rabi- and Dicke-type models, which are commonly used for the study of quantum light-matter interactions in cavity and circuit QED. We demonstrate that the validity of the two-level approximation, which is an essential step in this derivation, depends explicitly on the choice of gauge once the system enters the ultrastrong coupling regime. In particular, while in the electric dipole gauge the two-level approximation can be performed as long as the Rabi frequency remains much smaller than the energies of all higher-lying levels, it can dramatically fail in the Coulomb gauge, even for systems with an extremely anharmonic spectrum. We extensively investigate this phenomenon both in the single-dipole (Rabi) and multi-dipole (Dicke) case, and considering the specific examples of dipoles confined by double-well and by square-well potentials, and of circuit QED systems with flux qubits coupled to an LC resonator.
\end{abstract}
 
\maketitle

%
%


\maketitle
\section{Introduction}
In classical electrodynamics the invariance of Maxwell's equations under gauge transformations of the vector potential $\vec A$ and the  scalar potential $\Phi_{\rm el}$ is often used to simplify calculations by working in the most convenient gauge~\cite{Jackson}. In the formulation of the underlying theory of quantum electrodynamics (QED) the invariance of physical observables under local $U(1)$ gauge transformations is even taken as the fundamental ingredient from which QED is derived and the generalization of this principle to higher dimensional gauge-field theories forms the basis for modern particle physics. In atomic physics, quantum optics and solid-state physics, we are usually dealing with simplified models of QED to describe interactions between matter and electromagnetic fields. Although such models are based on various approximations, gauge invariance is in general still preserved. For example, the equivalence between the $\vec p\cdot \vec A$ interaction (Coulomb gauge) and the $\vec x\cdot \vec E$ interaction (electric dipole gauge) for evaluating resonant optical transition matrix elements for atoms is a common derivation found in many textbooks and introductory courses on quantum optics \cite{Scully,PhotonsAndAtoms}.

 There are, however, situations where the use of different gauges in quantum optical models is more subtle. For example, as first pointed out by Lamb~\cite{Lamb1952} and discussed further by others~\cite{Yang1976,Kobe1978,Lamb1987}, working in the Coulomb gauge or in the electric dipole gauge leads to slightly different predictions for a two-level atom driven by an off-resonant electric field. Related issues appear in the  evaluation of two-photon transition amplitudes, where depending on the choice of gauge, completely different sets of intermediate states must be considered to obtain converging results~\cite{Bassani1977}. 
The choice of gauge has also led to many controversies in the context of cavity QED, where the coupling of $N$ two-level atoms to a single radiation mode is frequently described by the Dicke model~\cite{Dicke1954,Brandes2005,Garraway2011}.  This model predicts a superradiant phase transition (SRT) \cite{Hepp1973,Wang1973}, when
the collective atom-field coupling reaches the ultrastrong coupling (USC) regime~\cite{Ciuti2005,FornDiaz2018}  and  becomes  comparable  to  the  optical and atomic frequencies. It was later shown---based on general sum-rule arguments---that this transition does not occur when the ``$A^2$-term" in the underlying minimal coupling Hamiltonian is  properly taken into account~\cite{Rzazewski1975}. However, by changing to the electric dipole gauge, this $A^2$-term can be eliminated~\cite{PhotonsAndAtoms,Keeling2007,Vukics2014,Griesser2016} and when restricted to a single mode, the original Dicke model---without any constraints on the coupling strength---can be recovered. This example shows that approximate models for light-matter interactions derived in different gauges may even lead to drastically different predictions, such as the existence or non-existence of a phase transition.

In a recent work \cite{Debernardis2018} it was shown that most of the ambiguities concerning the Dicke model and the superradiant phase transition can be fully resolved by a careful derivation and interpretation of the reduced effective cavity QED Hamiltonian. One of the important conclusions from this analysis was that the validity of the two-level approximation (TLA) for the dipoles depends explicitly on the choice of gauge, once the light-matter coupling becomes non-perturbative. 
Such conditions have been experimentally achieved in many solid-state implementations, using either collective excitations in dielectric materials \cite{Anappara2009,Todorov2010,Muravev2011,Schwartz2011,Geiser2012,Scalari2012,Benz2013,Dietze2013,
Scalari2013,KenaCohen2013,Mazzeo2014,Askenazi2014,Gubbin2014,Maissen2014,George2016,
Zhang2016,Bayer2017,Brodbeck2017,Askenazi2017}
or nonlinear elements in superconducting circuits \cite{Niemczyk2010,Forndiaz2010,Baust2016,Forndiaz2017,Yoshihara2017,Chen2017,Bosman2017,Gu2017}. 
The rich phenomenology which has been predicted to become observable in the USC regime has fuelled a remarkable research activity  in this domain \cite{Ashhab2010,Casanova2010,Braak2011,Hutchison2012,Ridolfo2012,Carusotto2012,Auer2012,Romero2012,Ashhab2013,DeLiberato2014,Ripoll2015,Bamba2015,
Hwang2015,Kockum2017,Garziano2017,Fedortchenko2016,Jaako2016,Cirio2016,LeBoite2016,Hagenmuller2016,
Armata2017,Beaudoin2011,Bamba2014,DeLiberato2017,Bamba2017,Bamba2017b,Flick2018}. It is thus fundamental to firmly establish under which conditions the usually-employed TLA is reliable or it can be made such by a proper choice of gauge.

In this work we provide such an analysis, which in particular illustrates the influence of the potential shape and of the number of dipoles on the validity of the TLA in the Coulomb and the electric dipole gauge. Remarkably, different results are obtained when considering single-dipole Rabi-type models, relevant for superconducting circuits, or multi-dipole Dicke and Hopfield models, which are instead usually employed to model dielectric systems. This remains true even for strongly anharmonic systems in which higher lying states can reasonably be considered out of resonance. In other words, in the USC regime, effective cavity-QED Hamiltonians, like the quantum Rabi or Dicke-type models, can only be consistently derived when the full system Hamiltonian is expressed in the appropriate gauge.

The remainder of the paper is structured as follows. In Sec. \ref{sec:The quantum Rabi model and the no-go theorem} we will first review the derivation of the quantum Rabi model and the resulting no-go- and counter-no-go theorems obtained in different gauges. This apparent contradiction is resolved in Sec. \ref{sec:Validity of the two-level approximation}, where we explicitly illustrate the invalidity of the TLA in the Coulomb gauge in terms of two specific examples. In Sec. \ref{sec:Multi-dipole cavity QED} we then extend these results to cavity QED systems with multiple dipoles. Finally, in Sec. \ref{sec:circuit_QED} we discuss the relevance of our findings in the context of  circuit QED and we conclude our work in Sec. \ref{sec:conclusions}.

\section{The quantum Rabi model and the no-go theorem}
\label{sec:The quantum Rabi model and the no-go theorem}
\begin{figure}
\centering
	\includegraphics[width=\columnwidth]{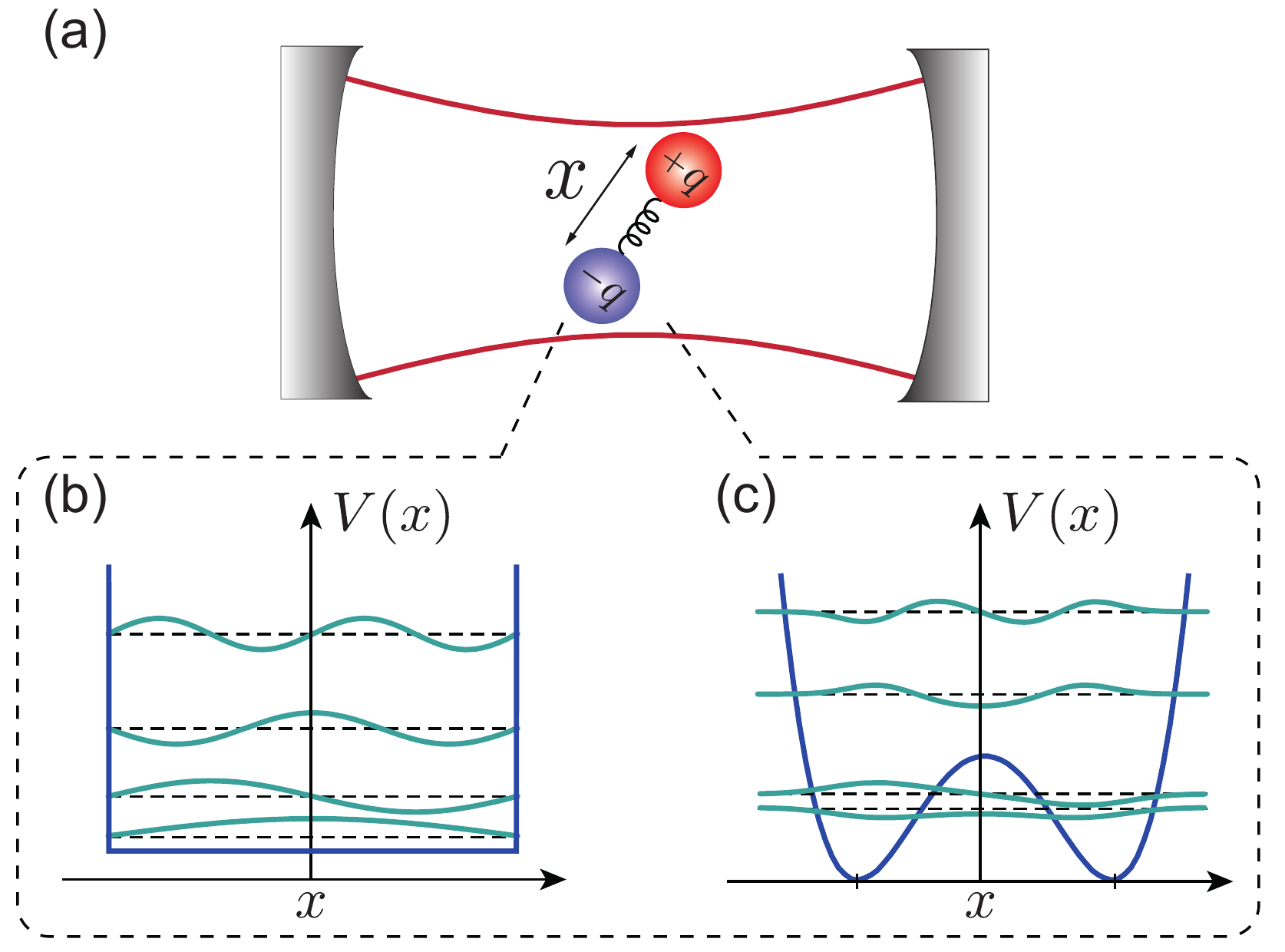}
	\caption{(a) Sketch of a generic cavity QED setup, where a dipole formed by two charges $+q$ and $-q$ is coupled to a single electromagnetic mode. The dipole is modeled as an effective particle of mass $m$ moving in a potential $V(x)$. Two prototype examples of an infinite square-well potential and a double-well potential are shown in (b) and (c), respectively.  In these plots, the dashed lines indicate the energies $E_n$ of the lowest bound states $|\varphi_n\rangle$ and the solid lines the shape of the corresponding wavefunctions.  }
	\label{fig:setup}
\end{figure}
For the following discussion we consider a generic setting as shown in Fig. \ref{fig:setup}, where a single electric dipole is coupled to a single mode of the electromagnetic field. The field mode is described by a harmonic oscillator with  bare frequency $\omega_c$ and annihilation (creation) operator $a$ ($a^\dag$).   Restricted to one dimension, the dipole can be modeled as an effective particle of mass $m$ in a potential $V(x)$, where $x$ is the separation between the charges $q$ and $-q$. Under these assumptions and making a dipole approximation, the Hamiltonian for this system is 
\begin{equation}\label{eq:ham_coulomb_generic_single}
H_{C} = \frac{\left(p - qA \right)^2}{2m} + V(x) + \hbar \omega_c a^\dag a,
\end{equation}
where $A=\mathcal{A}_0(a+a^\dag)$ is the vector potential along the $x$-direction with a zero-point amplitude $\mathcal{A}_0$. The form of $H_C$ follows directly from the  minimal coupling Hamiltonian, which is derived from the quantization of the electromagnetic field in the Coulomb gauge \cite{PhotonsAndAtoms}. Hamiltonian~\eqref{eq:ham_coulomb_generic} holds for any system interacting with a single electromagnetic mode via a dipole transition, for example, atoms, molecules, electrons in a quantum dot~\cite{Cottet2017}, etc.

\subsection{The quantum Rabi model in the Coulomb gauge}
\label{sec:rabi_model_coulomb}
By expanding the kinetic energy term, Hamiltonian~\eqref{eq:ham_coulomb_generic_single} can be divided into three contributions 
\begin{equation}\label{eq:ham_coulomb_generic_expanded}
H_C= H_d + \tilde H_{c} + H_{\rm int}^C.
\end{equation}
The first term, $H_d$, represents the bare Hamiltonian of the dipole, which can be diagonalized and written as
 \begin{equation}\label{eq:ham_bare_dipole}
H_d=  \frac{p^2}{2m} + V(x) = \sum_n  \hbar \omega_n |\varphi_n\rangle \langle \varphi_n|.
\end{equation}
Here $\omega_n$ is the eigenfrequency of the $n$-th motional eigenstate $|\varphi_n\rangle$. The second term in Eq. \eqref{eq:ham_coulomb_generic_expanded} represents the energy of the field mode including the $A^2$-term,
\begin{equation}
\tilde H_c= \hbar \omega_c a^\dag a +  \frac{q^2\mathcal{A}_0^2}{2m} (a+a^\dag)^2 = \hbar \tilde \omega_c c^\dag c.
\end{equation}
In the last step we have made a Bogoliubov transformation to express the field Hamiltonian in terms of new bosonic operators $c$ and $c^\dag$ and a renormalized frequency $\tilde \omega_c=\sqrt{\omega_c^2 +D^2}$, where $D^2= 2q^2\mathcal{A}_0^2\omega_c/(\hbar m)$. By making use of the relation $(a+a^\dag)= \sqrt{\omega_c/\tilde \omega_c}(c+c^\dag)$, the remaining dipole-field interaction term can be written as
\begin{equation}\label{eq:ham_coulomb_int_general}
H_{\rm int}^C= \frac{pA}{m} = \frac{q\mathcal{A}_0}{m} \sqrt{\frac{\omega_c}{\tilde \omega_c}} \sum_{n,k}  p_{nk}  (c+c^\dag)  |\varphi_n\rangle\langle \varphi_k|,
\end{equation}
where $p_{nk}= \langle \varphi_n| p|\varphi_k\rangle$ are the matrix elements of the momentum operator.

We are now interested in a simplified model for describing the near-resonant coupling of the dipole and the cavity mode, i.e., $\omega_c \approx \omega_{10}=\omega_{1}-\omega_0$, while all higher motional states are assumed to be far detuned. This can always be achieved for a sufficiently non-linear potential. Based on this assumption we make a TLA by restricting the sums in \eqref{eq:ham_bare_dipole} and \eqref{eq:ham_coulomb_int_general} to the lowest two states $\ket{\downarrow}\equiv  |\varphi_0\rangle$ and $\ket{\uparrow}\equiv -i |\varphi_1\rangle$. We then obtain the quantum Rabi model
\begin{equation}\label{eq:ham_rabi_coulomb}
H_{\rm Rabi}^C= \hbar \tilde \omega_c c^\dag c + \frac{\hbar g_C}{2} (c+c^\dag ) \sigma_x +  \frac{\hbar \omega_{10}}{2}\sigma_z,
\end{equation}
where  the $\sigma_k$ are the usual Pauli operators acting on the  subspace $\{\ket{\downarrow},\ket{\uparrow}\}$ and 
\begin{equation} 
g_C =\frac{2 q \mathcal{A}_0 \lvert p_{01}\rvert }{ \hbar m} \sqrt{\frac{\omega_c}{ \tilde \omega_c}},
\end{equation}
is the coupling strength in the Coulomb gauge.

\subsection{No-go theorem}
\label{sec:no_go_theorem}
In the USC regime, a central quantity of interest is the dimensionless coupling parameter
\begin{equation}\label{eq:zeta_C_def}
\zeta_C = \frac{g_C^2}{\tilde \omega_c \omega_{10}}.
\end{equation}
In the corresponding Dicke model for a large number  of $N\gg1$ dipoles, the value of $N \zeta^{(N)}_C= 1$ marks the onset of a ground state instability, i.e., the transition into a superradiant phase [see Sec. \ref{sec:Multi-dipole cavity QED}  below]. However, already for a single dipole a value of $\zeta_C \gtrsim 1$ results in a qualitative change in the ground state of the quantum Rabi model~\cite{Ashhab2010,Ashhab2013,Hwang2015}, which is associated with an exponential closing of the energy gap between the lowest two states and a large occupation of the photonic mode.

By using the general relation between the matrix elements of the position and the momentum operator,
\begin{equation}\label{eq:XPrelation}
p_{nk}= i m(\omega_{n}-\omega_k)  x_{nk},
\end{equation}
where $x_{nk}=\langle \varphi_n |x |\varphi_k\rangle$, this coupling parameter can be expressed as 
\begin{equation}\label{eq:NoGoTheorem}
\zeta_C =  \frac{D^2}{\omega_c^2+D^2} f  \leq   1,
\end{equation}
where we have introduced the oscillator strength 
\begin{equation}
f= \frac{2m \omega_{10}}{\hbar}|x_{10}|^2.
\end{equation}
For the last inequality in Eq.~\eqref{eq:NoGoTheorem} we have used the Thomas-Reiche-Kuhn (TRK) sum rule 
\begin{equation}
\sum_n   (\omega_n - \omega_0) |x_{n0}| ^2 = \frac{\hbar}{2m},
\end{equation} 
to place an upper bound on the value of  $f\leq 1$. This sum rule follows directly from $[x,[x, H_d]]=-\hbar^2/m$ and is valid for arbitrary potentials. Therefore, Eq.~\eqref{eq:NoGoTheorem} constrains the maximal value of the coupling strength in $H_{\rm Rabi}^C$, since by increasing the coupling, also the renormlaized cavity frequency $\tilde \omega_c$ increases accordingly. A similar calculation for $N$ dipoles  leads to an analogous constraint on the value of $N\zeta^{(N)}_C \leq 1$ \cite{Rzazewski1975,CiutiNatComm2010,ViehmannPRL2011}, which implies that the ground state of a cavity QED system always remains stable. Therefore, this bound is often called the `no-go theorem' for superradiant phase transitions.

\subsection{The quantum Rabi model in the dipole gauge}
Let us now repeat the derivation of the quantum Rabi model in the electric dipole gauge by first performing the unitary transformation $H_D = U H_C U^{\dag}$, where 
\begin{equation}
 U=\exp \left[  -i \frac{q x A}{\hbar} \right].
\end{equation}
In the dipole gauge we obtain 
\begin{equation}\label{eq:ham_dipolar_generic}
\begin{split}
H_{D} = \frac{p^2}{2m} + \tilde V(x) + \hbar \omega_c a^\dag a + i \omega_c q \mathcal{A}_0(a^{\dag}-a) x,
\end{split}
\end{equation}
where the potential $\tilde{V}(x)= V(x) + mD^2  x^2/2$ now includes an additional correction term from the coupling to the cavity field. As above, we diagonalize the Hamiltonian for the dipole,  
 \begin{equation}
\tilde H_d=  \frac{p^2}{2m} + \tilde V(x) = \sum_n  \hbar \tilde \omega_n |\tilde \varphi_n\rangle \langle \tilde \varphi_n|,
\end{equation}
and express the position operator in terms of the eigenstates $|\tilde \varphi_n\rangle$, i.e., $x=\sum_{n,k} \tilde x_{nk} |\tilde \varphi_n\rangle \langle \tilde \varphi_k|$, where $\tilde x_{nk}=\langle \tilde \varphi_n|x |\tilde \varphi_k\rangle$. Restricted to the two lowest states $\ket{\downarrow}\equiv  |\tilde \varphi_0\rangle$ and $\ket{\uparrow}\equiv |\tilde \varphi_1\rangle$ and introducing for convenience the rotated field operator $c=i a$, we  end up with the quantum Rabi Hamiltonian
\begin{equation}\label{eq:ham_dipolar_rabi}
H_{\rm Rabi}^D= \hbar\omega_c c^\dag c + \frac{\hbar g_D}{2}(c+c^\dag ) \sigma_x +  \frac{\hbar \tilde  \omega_{10}}{2}\sigma_z, 
\end{equation}
where 
\begin{equation}\label{eq:gD}
g_D = \frac{2\omega_c q \mathcal{A}_0 \lvert \tilde{x}_{10}\rvert}{\hbar}
\end{equation}
is the coupling strength in the dipole gauge. It depends on the matrix element $\tilde{x}_{10}$ between the two lowest eigenstates of the modified potential $\tilde{V}(x)$.

\subsection{Counter-no-go theorem}
\label{sec:counter_no_go}
Although $H_{\rm Rabi}^C$ and $H_{\rm Rabi}^D$ have exactly the same structure, the parameters that enter in the two models have a different dependence on the underlying system parameters. Therefore, it is interesting to  consider also the coupling parameter $\zeta_D=g_D^2/(\omega_c\tilde \omega_{10})$, which after some rearrangements can be expressed as 
\begin{equation}\label{eq:CounterNoGoTheorem}
\zeta_D = \frac{ D^2}{ \tilde {\omega}_{10}^2}\tilde f \leq \frac{D^2}{\tilde {\omega}_{10}^2}.
\end{equation}
In the last step we have again used the TRK sum rule for the bound on the oscillator strength $\tilde f=2m\tilde \omega_{10}|\tilde x_{10}|^2/\hbar \leq 1$. For a harmonically confined dipole, i.e., $V(x)=m\omega_{10}^2 x^2/2$, we find that $\tilde \omega_{10}^2=\omega_{10}^2+D^2$ and Eq.~\eqref{eq:CounterNoGoTheorem} reproduces the same bound as in Eq.~\eqref{eq:NoGoTheorem}. However, for an arbitrary potential there is \emph{a priori} no constraint on the ratio $D^2/\tilde \omega_{10}^2$. Therefore, in the dipole gauge the coupling parameter can in principle exceed this bound.
\begin{figure}
\centering
	\includegraphics[width=\columnwidth]{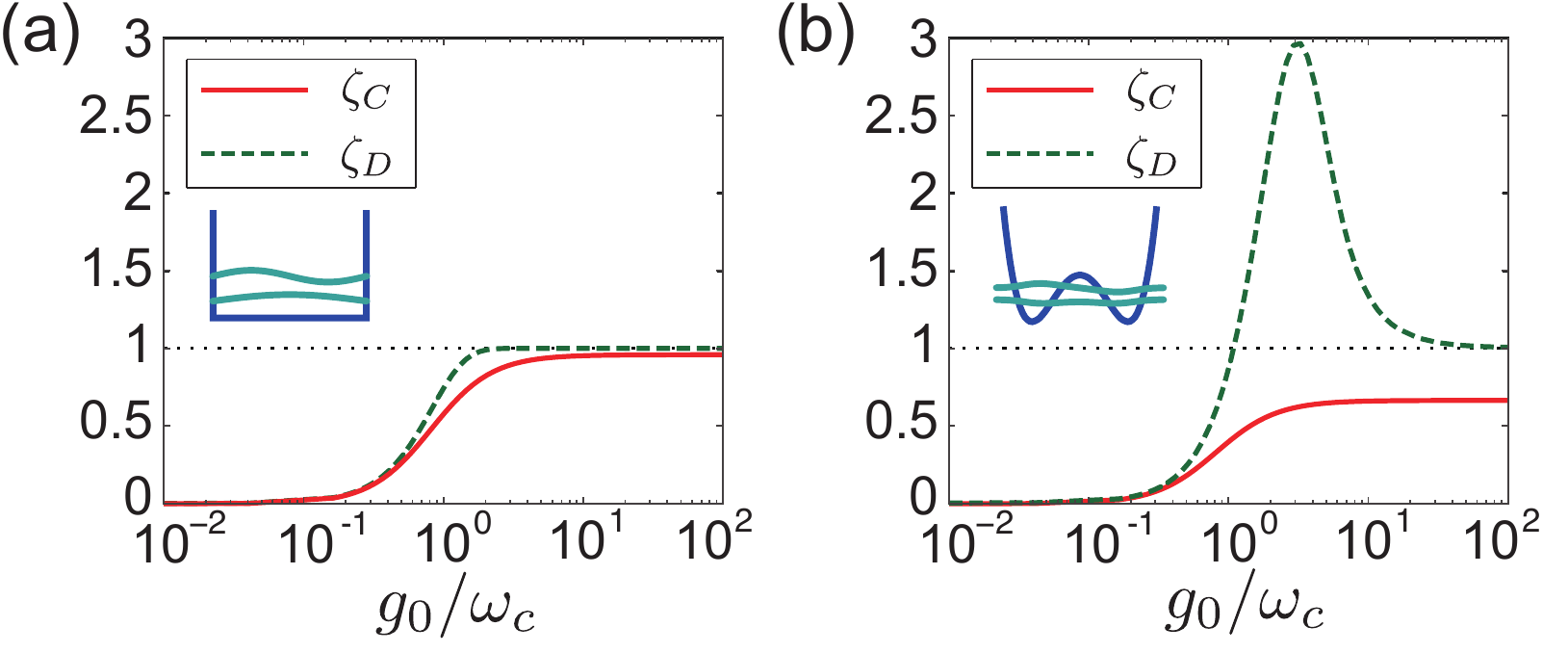}
	\caption{The dimensionless coupling parameters $\zeta_C$ and $\zeta_D$ as defined in Eq. \eqref{eq:zeta_C_def} and \eqref{eq:CounterNoGoTheorem} are plotted as a function of the bare  coupling strength $g_0\sim q$ for (a) a square-well potential and (b) a double-well potential with $\beta \approx 2.4$. 
	For both plots $\omega_c=\omega_{10}$ and the charge $q$ is used as a tunable parameter to vary the coupling strength. See Sec.~\ref{sec:Validity of the two-level approximation} for more details.}
	\label{fig:ZetaD_ZetaC_dw_sqw}
\end{figure}
To illustrate this point, we compare in Fig. \ref{fig:ZetaD_ZetaC_dw_sqw} the coupling parameters $\zeta_C$ and $\zeta_D$ for a square-well and a double-well potential. For this plot the charge $q$ is considered as a tunable parameter to vary the coupling strength, while all other system parameters are held fixed (see Sec.~\ref{sec:Validity of the two-level approximation} for more details).  We see that the two coupling parameters are indeed different. Most importantly, while in the case of a square-well potential both parameters remain below the value of one, in the case of a double-well potential $\zeta_D$ can considerably exceed this bound. 

\subsection{Role of the potential shape} 

The observed qualitative difference between different types of nonlinear potentials can be understood by focusing on the limit $q\rightarrow\infty$. In this limit, the correction term $\sim q^2x^2$ in the renormalized potential $\tilde V(x)$ dominates and localizes the eigenstates around $x=0$. Therefore, for any symmetric potential we can approximate
\begin{equation}
\lim_{q\rightarrow\infty} \tilde V(x)\simeq   \frac{m}{2} \tilde \omega_{10}^2  x^2,   \qquad \tilde \omega_{10}^2= (D^2+\Omega^2),
\end{equation}  
where 
\begin{equation}
\Omega^2=\frac{1}{m}
\left. \frac{\partial^2 V}{\partial x^2}\right|_{x=0},
\end{equation}  
is determined by the curvature of the potential at the origin. Since for a harmonic potential also the matrix element $|\tilde x_{10}|$ is maximized, we obtain
\begin{equation}
\lim_{q\rightarrow\infty} \zeta_D = \frac{D^2}{(D^2+\Omega^2)}.
\end{equation}
This shows that for a potential that is anti-confining at the origin, i.e., $\Omega^2<0$, the coupling parameter $\zeta_D$ approaches the value of one from above (see Appendix~\ref{app:Generalised counter no-go theorem} for a slightly more general derivation).
 This implies that  for such a potential a value of $\zeta_D>1$ can be achieved for a certain range of parameters, for which the no-go theorem does not hold.  From a purely classical point of view, the particle can lower its potential energy by moving from the center to one of the wells, which can compensate the electrostatic energy that is required to create a finite polarization.

\section{Validity of the two-level approximation}
\label{sec:Validity of the two-level approximation}
The discussion in the previous section shows that the quantum Rabi model $H_{\rm Rabi}^C$ derived in the Coulomb gauge and the corresponding model $H_{\rm Rabi}^D$ derived in the dipole gauge do not agree in general and can lead to qualitatively very different predictions. Since both models have been derived from the unitarily equivalent Hamiltonians $H_C$ and $H_D$, the TLA---which is the only approximation we made---must be invalid in at least one of the two gauges. In the following we explicitly illustrate this fact in terms of two concrete examples.

\subsection{Particle in a double-well potential}\label{sec:particle_in_dw}
\label{sec:subsec_Particle in a double-well potential}\label{sec:particle_in_dw}
As a first example we consider a dipole represented by a charged particle moving in a double-well potential, as depicted in Fig.~\ref{fig:setup}(c). In this case the Hamiltonian for the dipole is given by
\begin{equation}\label{eq:ham_sing_part_dw}
H_d = -\frac{\hbar^2}{2m}\frac{\partial^2}{\partial x^2} - \frac{\mu}{2}x^2 + \frac{\lambda}{4}x^4,
\end{equation}
where the two parameters $\mu,\lambda>0$ specify the shape of the double well. For the following discussion it is convenient to introduce the energy scale $E_d = \hbar^2/(mx_0^2)$ and the rescaled variable $\xi=x/x_0$, where $x_0=\sqrt[6]{ \hbar^2/(m\lambda)}$.  In terms of these quantities, Hamiltonian~\eqref{eq:ham_sing_part_dw} can be written as
\begin{equation}
H_d=  E_d \left( \frac{p_\xi^2}{2} - \frac{\beta}{2}\xi^2 + \frac{\xi^4}{4}\right),
\end{equation}
where $\beta = \mu m x_0^4/ \hbar^2$ and $p_\xi =-i \partial /\partial \xi$ is the dimensionless momentum operator. Similarly, the dipole-field interaction in the Coulomb gauge can be rewritten as 
\begin{equation}
H_{\rm int}^C=  \sqrt{\frac{\hbar D^2 E_d}{2\tilde \omega_c}} \sum_{n,k}  \langle  \varphi_n |p_\xi |  \varphi_k\rangle   (c+c^\dag)  |\varphi_n\rangle\langle \varphi_k|.
\end{equation}
In the dipole gauge we obtain  
\begin{equation}
\tilde H_d = E_d \left( \frac{p_\xi^2}{2} + \frac{(\gamma- \beta)}{2}\xi^2 + \frac{\xi^4}{4} \right),
\end{equation}
where $\gamma=\hbar^2 D^2/E_d^2$ accounts for the coupling-induced modification of the potential and 
\begin{equation}
H_{\rm int}^D=   \sqrt{\frac{\hbar^3 D^2 \omega_c }{2E_d} } \sum_{n,k}  \langle  \tilde \varphi_n |\xi |  \tilde \varphi_k\rangle   (c+c^\dag)  |\tilde \varphi_n\rangle\langle \tilde \varphi_k|,
\end{equation}
is the corresponding coupling Hamiltonian. In all numerical examples below, the value of $E_d$ will be fixed by the condition $\omega_c =\omega_1 - \omega_0$, which ensures that the bare cavity frequency is in resonance with the transition between the two lowest dipole levels in the limit of vanishing coupling.

\begin{figure}
\centering
	\includegraphics[width=\columnwidth]{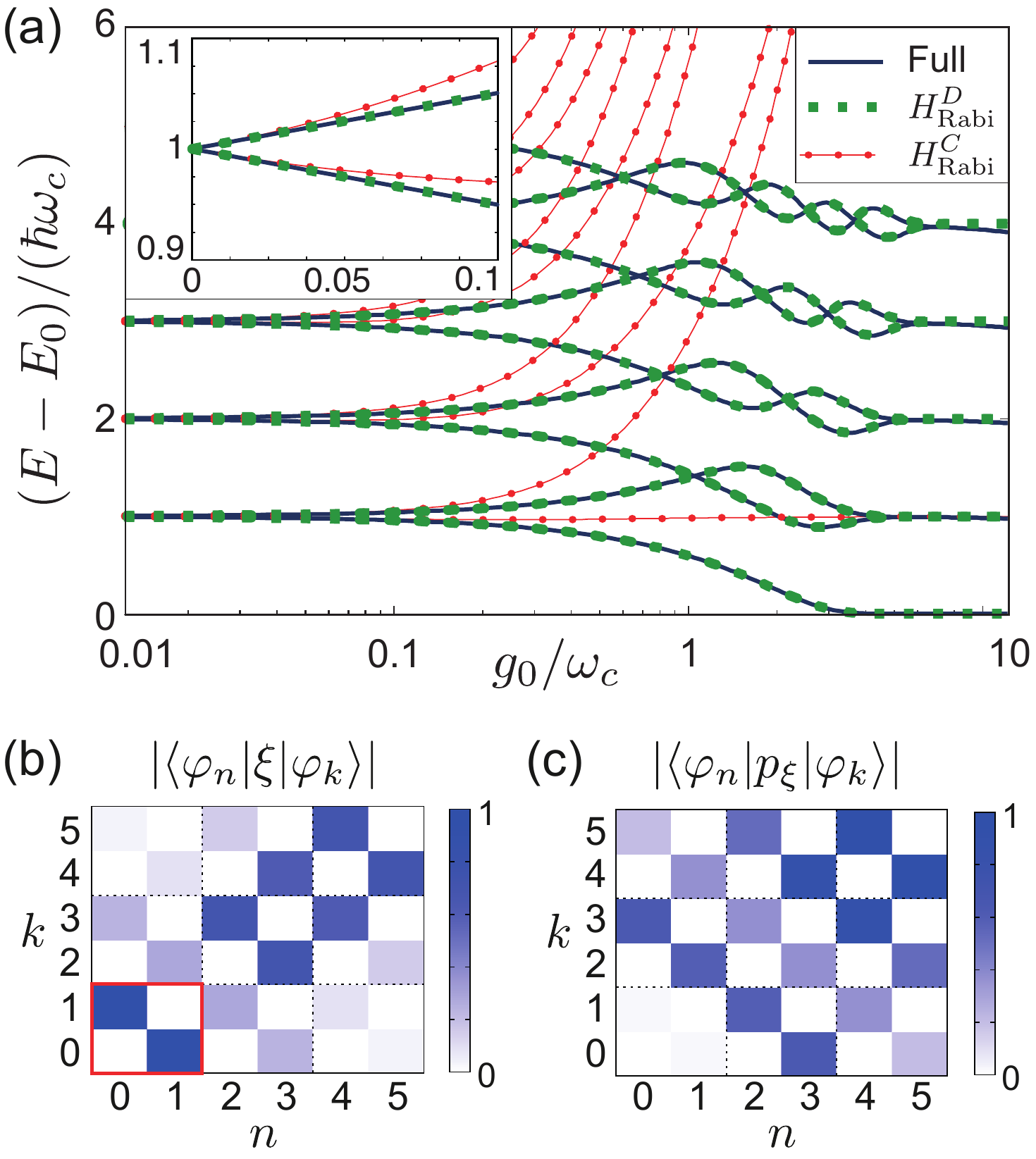}
	\caption{Double-well potential. (a) Comparison of the energy spectra obtained from the full model $H_C$ (solid blue line), the quantum Rabi model $H_{\rm Rabi}^D$ derived in the dipole gauge (green dashed line) and the quantum Rabi model $H_{\rm Rabi}^C$ derived in the Coulomb gauge (red dotted line). For these plots a double-well potential with parameters $\beta \approx 3.7$ and $\omega_{10} = \omega_c$ (which fixes the value of $E_d$) have been assumed. The inset shows a zoom of the predicted Rabi splitting between the first two excited energy levels. (b) Matrix elements of the dimensionless position operator $\xi$ and (c) matrix elements of the dimensionless momentum operator $p_\xi$ evaluated for the lowest eigenstates $|\varphi_n\rangle$ of the same double-well potential. 
For the sake of clarity, the values of the matrix elements have been normalized by the largest matrix element in each plot.}
	\label{fig:dw_combined}
\end{figure}

In Fig.~\ref{fig:dw_combined}(a) we plot the energies $E_n$ of the lowest eigenstates obtained from the reduced models $H_{\rm Rabi}^C$ and $H_{\rm Rabi}^D$ and compare these results with the exact eigenenergies obtained by diagonalizing the full Hamiltonian $H_C$. For this plot, $D\propto q\mathcal{A}_0$ is used as a tunable parameter to vary the coupling strength, while all other system parameters are held fixed. The resulting energies are then plotted as a function of 
\begin{equation}\label{eq:bare_coupling_g0}
\frac{g_0}{\omega_c}= \sqrt{\frac{2 \hbar D^2}{E_d \omega_c} } |\langle \varphi_0|\xi| \varphi_1\rangle|.
\end{equation}
Here $g_0$ denotes the bare coupling strength in the electric dipole gauge, i.e., the coupling defined in Eq.~\eqref{eq:gD}, but without taking any modification of the potential into account. 

The comparison in Fig.~\ref{fig:dw_combined}(a) shows that for very small values of the coupling, both models reproduce the expected vacuum Rabi splitting in the excited states, for example,  $(E_2-E_1)/\hbar \simeq g_D\simeq  g_C\simeq g_0$. However, already at moderate coupling strengths, $g_0/\omega_c\sim 0.1$, there are significant deviations in the predicted energies. More strikingly, for a value of $g_0/\omega_c=1$ the Rabi model in the Coulomb gauge already provides completely wrong predictions. This is very surprising, since for the chosen potential parameters, the frequency of the motional state $|\varphi_2\rangle$ is still very far detuned, i.e., $\Delta_{\rm nl}=(\omega_2-\omega_0)/(\omega_1-\omega_0) \approx 100$. Therefore,  from a naive estimate of the influence of higher motional states such a strong discrepancy is unexpected. For even larger values of $g_0/\omega_c\gtrsim 10$ (depending on the degree of nonlinearity) also the Rabi model $H_{\rm Rabi}^D$ becomes inaccurate and further corrections from the higher levels must be taken into account (see also Sec.~\ref{subsec:X2correction} below). 

\subsection{Origin of the break-down of the two-level approximation}

The observed break-down of the TLA in the Coulomb gauge at moderate couplings can be qualitatively understood~\cite{Debernardis2018} from relation~\eqref{eq:XPrelation}, which in terms of the normalized operators $\xi$ and $p_\xi$ reads
\begin{equation}
\langle  \varphi_n |p_\xi |  \varphi_k\rangle = i \hbar \frac{(\omega_{n}-\omega_k)}{E_d}  \langle  \varphi_n |\xi |  \varphi_k\rangle.
\end{equation}
This relation shows that the matrix elements of the momentum operator scale with the frequency difference between the coupled states. Therefore, transitions to energetically higher states are not systematically suppressed, since the large energy gap is compensated by a corresponding increase of the coupling matrix elements. 

This important difference between the position and the momentum operator is illustrated in more detail in Fig.~\ref{fig:dw_combined}(b) and (c), where the magnitudes of the matrix elements $\langle  \varphi_n|\xi| \varphi_k\rangle$ and $\langle  \varphi_n|p_\xi| \varphi_k\rangle$ are plotted for the lowest states of the double-well potential. We see that matrix elements of the position operator are always maximal between neighboring levels.  Therefore, transitions to energetically higher states are suppressed and for not too strong couplings we can restrict the dynamics of the dipole to the lowest two-level subspace. In contrast, for the momentum operator, the coupling to energetically higher states is much bigger than the coupling within the lowest two-level subspace and already for modest coupling strengths multiple levels must be taken into account to obtain an accurate description.  

This example shows that the difference between the Coulomb and the dipole gauge is rooted in the asymmetry between the position and the momentum operator. Such an asymmetry does not exist for the electromagnetic mode or for a harmonically bound dipole, where momentum and position operators are interchangeable. It is thus the nonlinearity of $V(x)$, which breaks this equivalence and favors the dipole gauge with an $x$-type coupling  for the purpose of deriving an effective two-level model.

\subsection{Particle in a square-well potential}\label{subsec:SW}
As a second example we consider a particle in an infinite square-well potential of width $L_w$ [see Fig.~\ref{fig:setup}(b)].  In this case we have $V(x)=0$ in the region $-L_w/2< x< L_w/2$ and $V(x)=\infty$ everywhere else. This potential mimics, for example, the transverse confinement of electrons in a semiconductor quantum well \cite{Ciuti2005}. For the square-well potential we define the characteristic length scale $x_0=L_w/2$ and the corresponding energy scale $E_d=\hbar^2/(mx_0^2)$. Otherwise we proceed as in Sec. \ref{sec:subsec_Particle in a double-well potential}.

\begin{figure}
\centering
	\includegraphics[width=\columnwidth]{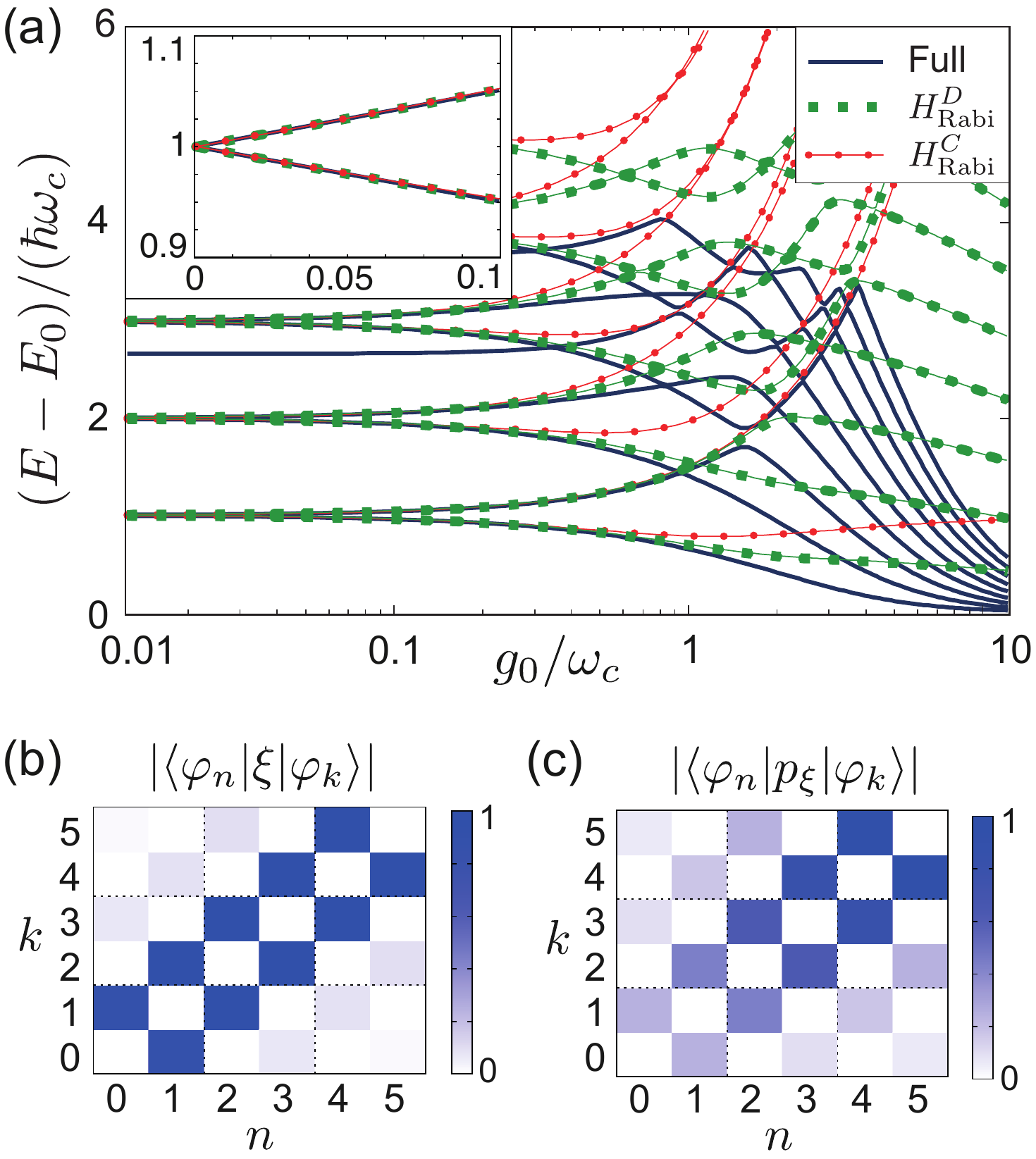}
	\caption{Square-well potential. (a) Comparison of the energy spectra obtained from the full model $H_C$ (solid blue line), the quantum Rabi model $H_{\rm Rabi}^D$ derived in the dipole gauge (green dashed line) and the quantum Rabi model $H_{\rm Rabi}^C$ derived in the Coulomb gauge (red dotted line). For these plots a square-well potential and $\omega_{10} = \omega_c$ (which fixes the value of $E_d$) has been assumed. The inset shows a zoom of the predicted Rabi splitting between the first two excited energy levels. (b) Matrix elements of the dimensionless position operator $\xi$ and (c) matrix elements of the dimensionless momentum operator $p_\xi$ evaluated for the lowest eigenstates $|\varphi_n\rangle$ of the same square-well potential. 
For the sake of clarity, the values of the matrix elements have been normalized by the largest matrix element in each plot.}
	\label{fig:sw_combined}
\end{figure}

Figure~\ref{fig:sw_combined}(a) shows the resulting comparison between eigenenergies $E_n$ obtained from the two Rabi models $H_{\rm Rabi}^C$ and $H_{\rm Rabi}^D$ and the full model $H_C$. Overall we see a very similar trend as for the double-well potential. The energy levels obtained from $H_{\rm Rabi}^C$ show significant deviations from the exact energies for $g_0/\omega_c\gtrsim 1$. 
For the square-well potential also the predictions of $H_{\rm Rabi}^D$ are rather poor for only slightly higher couplings $g_0/\omega_c\gtrsim 2$. This is related to the fact that for a square-well potential  the degree of nonlinearity, $\Delta_{\rm nl}=8/3$, is fixed and much smaller than for the double-well potential considered above.  Note that for the square-well potential the exact energies do not exhibit an exponentially suppressed energy gap for large couplings. The spectrum rather becomes more and more harmonic with a vanishing frequency for $g_0/\omega_c\gg 1$. This is expected from a model of two coupled oscillators in the limit where the dipole potential is dominated by the correction term, i.e., $\tilde V(x)\approx mD^2 x^2/2$.

Figure~\ref{fig:sw_combined}(b) and (c) show again the matrix elements of the position and the momentum operator, which are now evaluated for the lowest states of the square-well potential. We see that the structure of the matrix elements is more similar and already closer to that of a harmonic potential. Although in the dipole gauge the anharmonicity in the energy spectrum still allows us to identify a isolated two-level subspace for values of $g_0/\omega_c\leq 1$, this is no longer possible for slightly higher couplings.

\subsection{The Rabi splitting and the oscillator strength}

By comparing the Rabi splittings in the insets of Fig.~\ref{fig:dw_combined}(a) and Fig.~\ref{fig:sw_combined}(a) for small and moderate values of $g_0/\omega_c\lesssim 0.1$, we find that the deviation of the spectrum of $H_{\rm Rabi}^C$ is less significant for the square-well potential than for the double-well potential. To understand this difference we calculate the energies $E_n$ of $H_{\rm Rabi}^C$ and $H_{\rm Rabi}^D$ up to second order in $g_0/\omega_c$. For $\omega_c=\omega_{10}$ we obtain
\begin{equation}
\Delta E^C_{1,2}=E^C_{1,2} - E^C_0\simeq  \hbar \omega_{c} \mp \frac{\hbar g_0}{2}+  \frac{\hbar g_0^2}{4\omega_c}\frac{1}{f}  ,
\end{equation}
for the first two excitation energies in the Coulomb gauge and
\begin{equation}
\Delta E^D_{1,2}=E^D_{1,2}-E^D_0\simeq  \frac{\hbar \omega_c + \hbar \tilde{\omega}_{10}}{2} \mp \frac{\hbar g_0}{2},
\end{equation}
in the electric dipole gauge.  
While up to second order in $g_0/\omega_c$ the predicted Rabi splitting, $(E_2-E_1)/\hbar\simeq g_0$ still agrees in both gauges, we observe a systematic blue shift of the energy levels in the Coulomb gauge. For a sufficiently non-linear potential where $\tilde{\omega}_{10}\simeq \omega_{10}$, this artificial blue shift reads
\begin{equation}\label{eq:JC_discrepancy_C_D_lowest_order}
\Delta E^C_1 - \Delta E^D_1  \simeq \frac{\hbar g_0^2}{4\omega_c}\frac{1}{f} .
\end{equation}
This result immediately explains the strong discrepancies between the energy levels observed in the inset of Fig.~\ref{fig:dw_combined}(a) for the double-well potential. In this case $f_{\rm dw}\simeq 0.1$ and therefore an appreciable deviation of the predicted energy levels occurs already  at the onset of the USC regime. Instead, for the square-well potential, where $f_{\rm sq}\approx 0.96$, the differences between $H_{\rm Rabi}^C$ and $H_{\rm Rabi}^D$ become significant only at larger couplings. 

This comparison shows that apart from the degree of nonlinearity $\Delta_{\rm nl}$ and the curvature of the potential at the origin, $\Omega^2$,  the oscillator strength $f$ of the lowest dipole transition is a third characteristic parameter, which affects the validity or non-validity of the TLA. In particular, this parameter determines the validity of the TLA in the Coulomb gauge at moderate interaction strengths. Given the upper bound $f\leq 1$, a value of $f\approx 1$ means that the coupling of the ground state to states $|\varphi_{n\geq 2}\rangle$ are suppressed by vanishingly small coupling matrix elements for both the position and the momentum operator.

\section{Multi-dipole cavity QED}
\label{sec:Multi-dipole cavity QED}

The USC coupling regime was observed for the first time exploiting collective electronic transitions between the subbands of doped quantum wells \cite{Anappara2009}, and many-electron dielectric systems remain today one of the leading platforms for USC physics. This is due to the large density of dipoles achievable, which translates into a large collective coupling strength. Moreover, as already mentioned in the introduction, the interest in the bound for the coupling parameter $\zeta_C$ originally emerged from debates over the existence or non-existence of the superradiant phase transition in cavity QED systems with a large number of dipoles.
It is thus of paramount importance to extend the previous investigation on gauge non-invariance to the case of multi-dipole cavity QED. 

By assuming for simplicity a homogeneous mode function for the electromagnetic field, we can model a multi-dipole system with the minimal coupling Hamiltonian 
\begin{equation}\label{eq:ham_coulomb_generic}
H_{C} = \sum_{i=1}^N \left[ \frac{\left(p_i - qA \right)^2}{2m} + V(x_i)\right] + \hbar \omega_c a^\dag a +H_{\rm dd}.
\end{equation}
Here the last term, $H_{\rm dd}$, accounts for direct dipole-dipole interactions, which depend in detail on the precise arrangement of the dipoles and the geometry of the setup. Since the form of $H_{\rm dd}\sim x_i x_j$ is invariant under the gauge transformation used below, it does not directly affect the following arguments about the difference between the Coulomb and the dipole gauge. Therefore, in the remainder of this section we will simply omit this term and refer the reader to Ref.~\cite{Debernardis2018} for a more detailed discussion about dipole-dipole interactions in single-mode cavity QED systems.

\subsection{The Dicke model in the Coulomb gauge}
By proceeding the same way as in Sec.~\ref{sec:The quantum Rabi model and the no-go theorem}, we perform a TLA for each of the dipoles and  readily obtain the Dicke model 
\begin{equation}\label{eq:ham_dicke_coulomb}
H_{\rm DM} = \hbar \tilde \omega_c c^{\dag} c + \hbar \omega_{10} S_z + \hbar g_C (c + c^{\dag})S_x,
\end{equation}
where $S_k=1/2\sum_{i} \sigma_k^i$ are collective spin operators. In Eq.~\eqref{eq:ham_dicke_coulomb} all the parameters are the same as in $H_{\rm Rabi}^C$ in Eq.~\eqref{eq:ham_rabi_coulomb}, except that the cavity frequency $\tilde \omega_c\equiv \tilde \omega_c(N)=\sqrt{\omega_c^2 + N D^2}$ is now renormalized by the presence of $N$ dipoles.  In the limit $g_C\rightarrow 0$ the ground state of $H_{\rm DM}$ is the normal vacuum state with all dipoles in state $\ket{\downarrow}$ and the photon mode in state $|0_c\rangle$. For a large number of dipoles, $N\gg1$, we can then use a Holstein-Primakoff transformation \cite{HolsteinPrimakoff} to evaluate the frequencies $\omega_{\pm}$ of the two collective polariton modes, 
\begin{equation}\label{eq:HP_polariton_modes_coulomb}
\omega_{C\pm}^2 =\frac{1}{2}\left[\omega_{10}^2 + \tilde\omega_c^2 
\pm \sqrt{( \tilde \omega_c^2- \omega_{10}^2 )^2 + 4 N g_C^2 \tilde \omega_c\omega_{10} }\right].
\end{equation}
When $N g_C^2
> \tilde \omega_c \omega_{10}$ the lower polariton mode becomes unstable, i.e., $\omega_-^2<0$, and a transition into a superradiant phase occurs. However, similar to the bound derived in Eq.~\eqref{eq:NoGoTheorem} we obtain~\cite{Rzazewski1975,ViehmannPRL2011}
\begin{equation}
N \zeta_C^{(N)}=\frac{N g_C^2}{\tilde \omega_c(N) \omega_{10}} \leq  \frac{ND^2}{\omega_c^2 + ND^2}< 1,
\end{equation}
which implies that this phase transition point cannot be reached.

\subsection{The extended Dicke model in the dipole gauge}
\label{sec:EDM_in_dip_gauge}
We can repeat the same derivation in the dipole gauge, starting from the Hamiltonian $H_D=UH_C U^\dag$, where  $U=\exp(-i q A \sum_i x_i/\hbar)$. After this transformation we obtain 
\begin{equation}\label{eq:ham_many_dip_dipgauge}
\begin{split}
H_{D} = & \sum_i \left[ \frac{p_i^2}{2m} + \tilde V(x_i) \right]  + \frac{mD^2}{2}\sum_{i\neq j} x_i x_j  \\
&+ \hbar \omega_c a^\dag a + i \omega_c q \mathcal{A}_0(a^{\dag}-a) \sum_i x_i.
\end{split}
\end{equation}
We see that apart from the corrections to the confining potential $\tilde V(x_i)=V(x_i)+mD^2 x_i^2/2$,  already encountered for a single dipole, Hamiltonian $H_{D}$ now contains additional interactions $\sim x_i x_j$ between the dipoles. 
Therefore, after performing the TLA and setting $c=i a$, we obtain the extended Dicke model~\cite{Jaako2016,Debernardis2018}
\begin{equation}\label{eq:ham_EDM}
H_{\rm EDM} = \hbar \omega_c c^{\dag} c + \hbar \tilde \omega_{10} S_z + \hbar g_D (c + c^{\dag})S_x + \frac{\hbar g_D^2}{\omega_c}S_x^2,
\end{equation}
which is no longer of the same form as Hamiltonian $H_{\rm DM}$ derived in the Coulomb gauge. It contains an additional all-to-all interaction term, which corresponds to the so-called ``$P^2$-term" in the electric dipole gauge Hamiltonian~\cite{PhotonsAndAtoms,Todorov2010,Todorov2012,Todorov14}.

Similar to the case of the Dicke model, we can analyze the stability of the ground state of $H_{\rm EDM}$ by using a Holstein-Primakoff transformation in the limit $N \rightarrow \infty$, but keeping $\sqrt{N}g_0$ finite. For the resulting polariton frequencies we obtain 
\begin{equation}\label{eq:HP_polariton_modes_dipolar}
\omega_{D\pm}^2 = \frac{1}{2}\left[\Omega_{10}^2 + \omega_c^2 \pm \sqrt{\left( \Omega_{10}^2 - \omega_c^2 \right)^2 + 4 N g_D^2 \tilde \omega_{10}\omega_c} \right],
\end{equation}
where $\Omega_{10}=\sqrt{ \tilde \omega_{10}( \tilde \omega_{10}+Ng_D^2/\omega_c)}$. The condition for an unstable mode is now given by $Ng^2_D>\Omega_{10}^2 (\omega_c/\tilde \omega_{10})$. However, after expressing $g_D^2=(2\omega_c m |\tilde x_{10}|^2/\hbar)D^2$ and using the TRK sum rule, we obtain the bound 
\begin{equation}
\frac{N g_D^2 }{\Omega_{10}^2 (\omega_c/ \tilde \omega_{10})}\leq    \frac{ND^2}{\tilde \omega_{10}^2+ ND^2}< 1,
\end{equation}
showing that also in the dipole gauge no instability occurs~\cite{Jaako2016,Todorov2012}. Although in this case the single-dipole coupling is not constrained by any bound, the inclusion of the $S_x^2$ term stabilizes the system for $N\gg 1$. We emphasize that this no-go-theorem holds for $H_{\rm dd}=0$, where the dipoles are only coupled to a single cavity mode, but not directly among each other. If direct dipole-dipole interactions are included, there can be additional ferroelectric instabilities (in both gauges)~\cite{Keeling2007,Griesser2016}, which, however, occur only for very specific geometries~\cite{Debernardis2018}. 

\subsection{Polariton spectra and fake depolarization shifts}
\label{sec:fake_dep_shift}

\begin{figure}
\centering
	\includegraphics[width=\columnwidth]{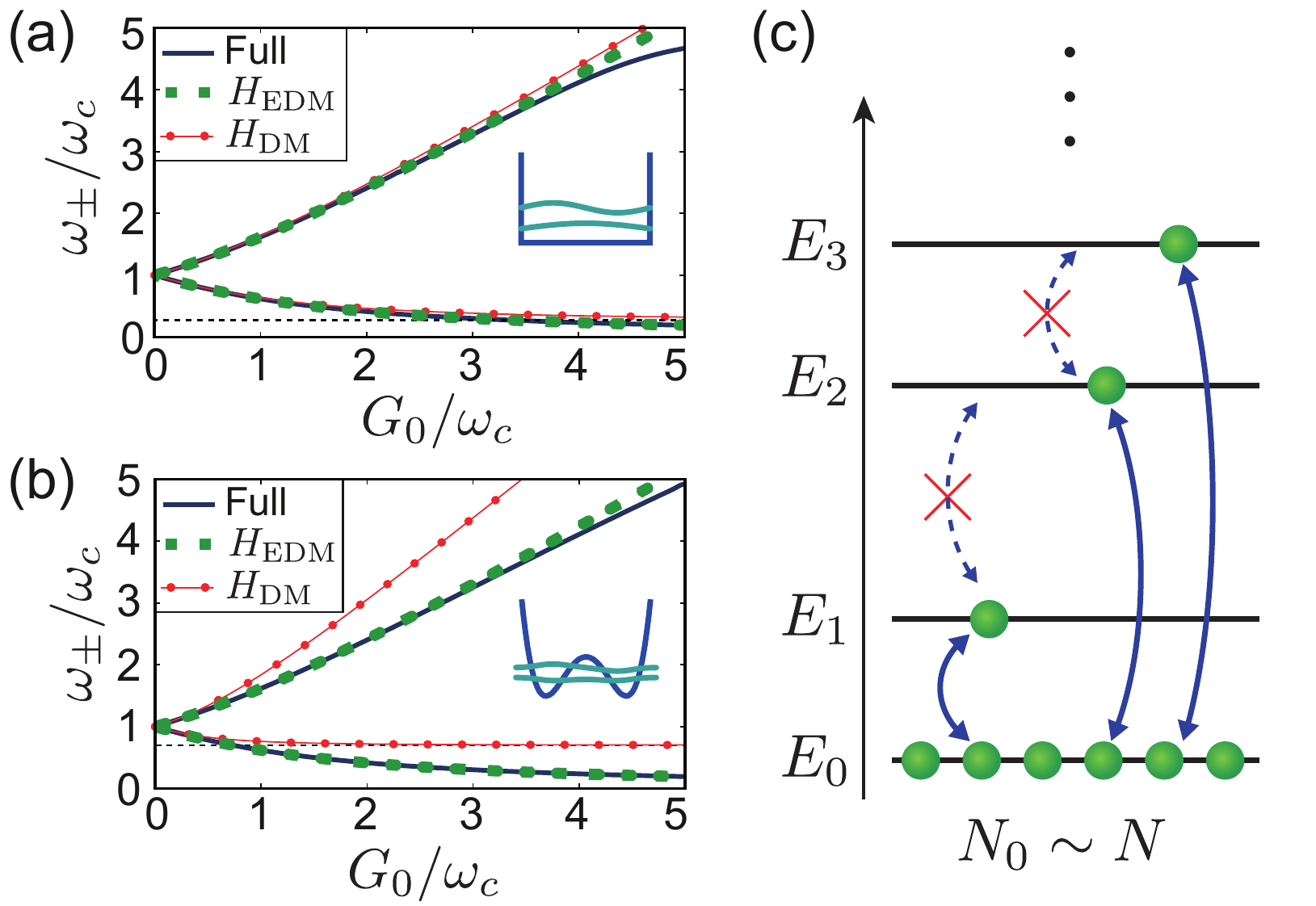}
	\caption{The frequencies $\omega_\pm$  of the two lowest polariton modes  are plotted in the limit  $N\gg 1$  as a function of the collective coupling strength $G_0=g_0\sqrt{N}$ and for $\omega_{10} = \omega_c$. In (a) the case of a square-well potential ($f_{\rm sq}\approx 0.96$) and in (b) the case of a double-well potential with $\beta \approx 2.3$,  and $f_{\rm dw}\approx 0.71$ is considered. The different lines represent the results obtained from the Dicke model $H_{\rm DM}$ derived in the Coulomb gauge  ($\omega_{C\pm}$, red solid-dotted), the extended Dicke model $H_{\rm EDM}$ derived in the electric dipole gauge ($\omega_{D\pm}$, green squares) and the two lowest branches of the full spectrum (blue solid line). In both plots, the horizontal dashed line represents the fake depolarization shift in the Coulomb gauge, as given in Eq.~\eqref{eq:DepolarizationShift}. (c) Sketch of the relevant energy levels of the full multi-dipole Hamiltonian $H_C$ in the weak-excitation regime. In this limit, most dipoles occupy the lowest potential state with energy $E_0$  and cavity-induced transitions between pairs of higher states can be neglected.  }
	\label{fig:HP_polariton_modes}
\end{figure}

Although for $N\gg 1$ there is a qualitative agreement on the stability of the system, it is important to keep in mind that the spectra given in Eq.~\eqref{eq:HP_polariton_modes_coulomb} and Eq.~\eqref{eq:HP_polariton_modes_dipolar} are in general not identical. This is illustrated in Fig.~\ref{fig:HP_polariton_modes}(a) and (b), where we compare the polariton frequencies  $\omega_{C\pm}$ and $\omega_{D\pm}$ for the two cases of a square-well and a double-well potential. In the dilute regime, in which the number of excitations is much smaller than the number of dipoles $N$, also the full Hamiltonian $H_C$ in Eq. \eqref{eq:ham_coulomb_generic} can be solved by bosonizing the matter excitations using the Holstein-Primakoff transformation or one of other essentially equivalent techniques \cite{HolsteinPrimakoff,Ciuti2005,Combescot2008}. 
These transform $H_C$ in a quadratic, bosonic Hamiltonian, which can be easily diagonalized (see Appendix~\ref{app:no-go_dep_shift}). In Fig. \ref{fig:HP_polariton_modes}(a) and (b) the solid lines represent the resulting exact polariton frequencies.   

Note that physically the bosonization of a collection of dipoles is justified by the fact that the probability of a photon to be absorbed by a single dipole scales as $1/N$, and saturation effects vanish [see Fig.~\ref{fig:HP_polariton_modes}(c)]. Ladder transitions, which would couple, for example, the first excited state $\ket{\varphi_1}$ to higher lying states $\ket{\varphi_{n>1}}$ are negligible for $N\gg 1$. We can thus expect the TLA, which neglects all transitions between the first two states and the higher excites ones, to be a better approximation in the multi-dipole case. In particular, it has to become exact for a harmonic confinement potential, as in this case $f=1$ and also all transition matrix elements between the ground state $\ket{\varphi_0}$ and all the excited states $\ket{\varphi_{n>1}}$ vanish according to the TRK sum rule.

Figure \ref{fig:HP_polariton_modes} shows that also in the collective, weak-excitation regime the agreement between the predictions from the Dicke model and the extended Dicke model depend  on the shape of the dipole potential. To see this dependence more explicitly, we write  $g_D^2\simeq g_0^2=(\omega_c/\omega_{10})f D^2$ and $g_C^2=(\omega_{10}/\tilde \omega_c)f D^2$, where we have assumed $\tilde \omega_{10}\simeq \omega_{10}$ and $\tilde x_{10}\simeq x_{10}$. This approximation is justified for $N\gg1$, where $g_0\sim 1/\sqrt{N}$ is small and corrections to the potential of a single dipole can be neglected. After some rearrangements we obtain 
\begin{equation}
\begin{split}
&\omega_{C\pm}^2 = \frac{1}{2}\left[\omega_{10}^2 + \omega_c^2+ ND^2 \pm \right.\\
&\left.  \sqrt{\left(\omega_{10}^2 + \omega_c^2+ ND^2  \right)^2 -  4 \omega_c^2 \omega_{10}^2- 4N(1-f)D^2 \omega_{10}^2} \right],
\end{split}
\end{equation}
in the Coulomb gauge and 
\begin{equation}
\begin{split}
\omega_{D\pm}^2 =& \frac{1}{2}\left[\omega_{10}^2 + \omega_c^2+ f ND^2 \pm \right.\\ &\left. \sqrt{\left(\omega_{10}^2 + \omega_c^2+ f ND^2  \right)^2 -  4 \omega_c^2 \omega_{10}^2} \right],
\end{split}
\end{equation}
in the dipole gauge. 

As expected from the general argument above, for harmonically confined dipoles the TLA spectra obtained in the Coulomb gauge and in the dipole gauge are identical  to the exact one. But also for a square-well potential ($f_{\rm sq} \approx 0.96$) as relevant for intersubband polaritons \cite{Ciuti2005}, there is no significant difference.   However, for general potentials the oscillator strength $f$ can be much smaller than one and a notable discrepancy between the spectra can occur. 
This is illustrated in Fig.~\ref{fig:HP_polariton_modes}(b) for the example of a double-well potential with $f \approx 0.7$. Specifically,  on resonance,  $\omega_{10}=\omega_c$, and up to lowest order in the collective coupling $G_0=g_0\sqrt{N}$, we find
\begin{equation}
\omega_{C\pm} - \omega_{D\pm} \simeq \frac{G^2_0}{4\omega_c}\left(\frac{1}{f} - 1 \right).
\end{equation}
Therefore, similar to the result obtained for a single dipole in Eq.~\eqref{eq:JC_discrepancy_C_D_lowest_order}, the difference  disappears in the case of harmonic dipoles or for potentials with an almost saturated oscillator strength. However, for all other potentials the excitation spectra obtained for effective Hamiltonians in different gauges can exhibit significant deviations once the \emph{collective} USC regime, $G_0\sim \omega_c$, is reached.

In general the Dicke model derived in the Coulomb gauge predicts a blue shift of the spectrum and a finite frequency at large couplings
\begin{equation}\label{eq:DepolarizationShift}
\lim_{G_0 \rightarrow \infty} \omega_{C -} = \omega_{10}\sqrt{1-f
} >0.
\end{equation} 
 Such a depolarization shift of the spectrum is in principle expected from the additional effect of dipole-dipole interactions $\sim H_{\rm dd}$, which are, however, explicitly omitted in the present analysis.  Therefore, this apparent depolarization shift is a pure artifact of the TLA and disappears when more and more levels are included. In contrast, in the dipole gauge, including the lowest two levels is already a very good approximation, up to very large values of the collective coupling $G_0$.

\subsection{
Discussion: Cavity QED}

In solid-state cavity QED, the two platforms in which USC with the largest couplings has been observed, are Landau polaritons and intersubband polaritons. Landau polaritons have, to the best of our knowledge, been theoretically investigated using only the Coulomb gauge \cite{Hagenmuller10,Hagenmuller12,Pellegrino14}. They presently hold the absolute world record for the observed normalised coupling, with a measured value of  $G/\omega_c\approx 2.86$ \cite{Bayer2017}. Notwithstanding such large values, our analysis shows that the two gauges are still equivalent in this system, because the electrons are confined by the perfectly harmonic potential due to the magnetic field.

Instead, in the case of intersubband polaritons, both theories based on the Coulomb~\cite{Ciuti2005,DeLiberato2012} and on the electric dipole~\cite{Todorov2010,Todorov2012,Todorov2015} gauges have been used. The two approaches led to slightly different predictions, which, however, cannot  be trivially interpreted in the light of the present results due to their more microscopic nature, which includes the intrinsically multi-mode nature of the photonic cavity and  a different treatment of dipole-dipole interactions. Experimentally, all investigations of intersubband polaritons in the USC regime have been performed using either rectangular \cite{Anappara2009,Askenazi2014} or parabolic \cite{Geiser2012} quantum wells. Considering that the record normalised coupling achieved in intersubband polaritons is $G/\omega_c\approx 0.9$ \cite{Askenazi2017}, we would thus expect that theories based on both gauges provide a quantitatively correct fit of existing data. The use of intersubband polaritons in  asymmetric quantum wells has been proposed to achieve terahertz inter-polariton emission, with the possibility to engineer the values of dipoles between  different states \cite{DeLiberato2013}.
The previous results show that an extension of such proposals to the USC regime, required for describing emission in the mid-infrared range, would work only in the dipolar gauge, or without performing the TLA and considering instead the full set of electronic states.

\section{Few-dipole USC and gauge non-invariance in circuit-QED}
\label{sec:circuit_QED}

The stability of the ground state predicted by Eqs.~\eqref{eq:HP_polariton_modes_coulomb} and~\eqref{eq:HP_polariton_modes_dipolar} for both gauges seems to contradict the findings from Secs.~\ref{sec:The quantum Rabi model and the no-go theorem}  and~\ref{sec:Validity of the two-level approximation}, where in the dipole gauge even for a single dipole an exponential closing of the energy gap, i.e., a precursor of a phase transition, was found. Here it is important to keep in mind that the results in Eqs. \eqref{eq:HP_polariton_modes_coulomb} and \eqref{eq:HP_polariton_modes_dipolar} have been derived in the limit $N\rightarrow \infty$. By keeping the resulting collective coupling $G_0=g_0\sqrt{N}$ finite, taking this limit also implies $g_0\rightarrow 0$. To complete our comparison of the two gauges,  it is thus necessary to consider also the intermediate regime, where $N>1$ and $g_0/\omega_c\sim 1$, and nonlinear, few-body and USC effects play a role. As can be seen from Eqs.~\eqref{eq:ham_dicke_coulomb} and~\eqref{eq:ham_EDM}, these effects are described in the Coulomb gauge and in the electric dipole gauge by two different effective models. 

Although dielectric platforms are progressively approaching the regime of few-electron USC~\cite{Todorov14,Keller17}, for the moment the condition $g_0/\omega_c\sim 1$ is accessible only in circuit QED, where superconducting qubits can be coupled very strongly to microwave resonators. Therefore, in this section we will  explicitly focus on a circuit QED setup with flux qubits, where the USC regime with individual qubits has already been demonstrated~\cite{Niemczyk2010,Forndiaz2010,Baust2016,Forndiaz2017,Yoshihara2017,Chen2017}. \emph{A priori} it might not be obvious that the effective models describing such macroscopic circuits should be directly related to the microscopic QED Hamiltonians discussed in the previous sections.  In particular, for flux-based qubits both the dynamical variables and the physical coupling mechanism are very different from the scenario investigated  above. However, as we will now show for a specific example, the structure of the Hamiltonians that appear in the description of circuits is often very similar to regular cavity QED, meaning that also the choice of gauge becomes a relevant issue.

\subsection{Circuit QED}

\begin{figure}
	\includegraphics[width=\columnwidth]{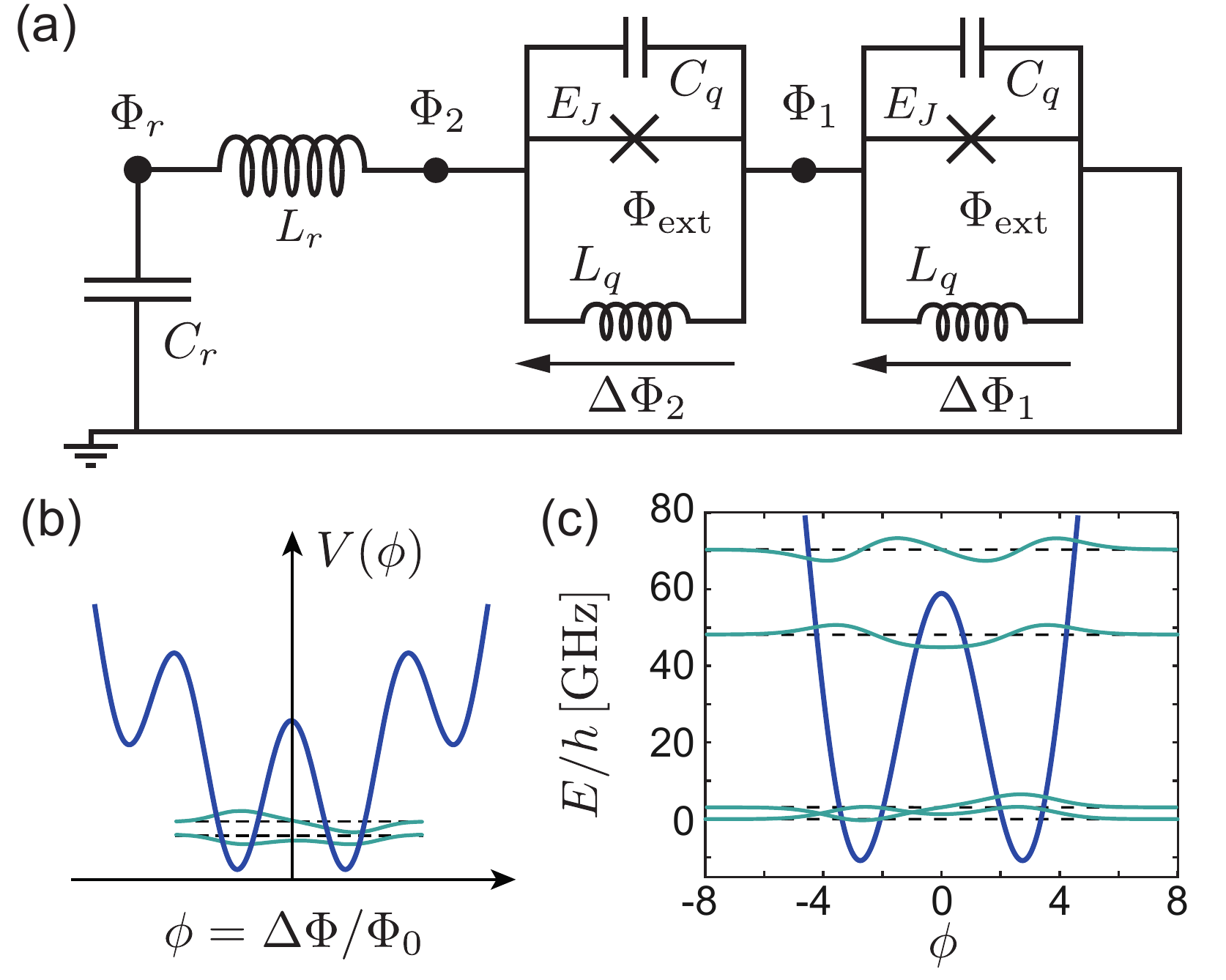}
	\caption{Circuit QED with flux qubits. (a) Sketch of a multi-qubit circuit QED system, where two flux qubits are coupled to a lumped-element $LC$ resonator with inductance $L_r$ and capacitance $C_r$. In the simplest case, each flux qubit is realized by an rf-SQUID circuit and can be modeled as an effective particle with a dimensionless coordinate $\phi=\Delta \Phi/\Phi_0$ moving in an effective potential $V(\phi)$. (b) Typical shape of the potential $V(\phi)$ for a generic flux-qubit
	 where the two lowest tunnel-coupled states form an isolated two-level subspace.
	(c) Shape of the potential $V(\phi)$ and the lowest eigenstates $|\varphi_n\rangle$ for a specific flux qubit with parameters $E_{L_q}/h = 7$ GHz, $E_{C_q}/h =12$ GHz and $E_J/h= 50$ GHz. 
	}
	\label{fig:circuit}
\end{figure}

Figure~\ref{fig:circuit} shows a prototype circuit QED setup~\cite{Jaako2016,Armata2017} with $N=2$ superconducting flux qubits coupled in series to a lumped-element $LC$ resonator with inductance $L_r$ and capacitance $C_r$. 
Following the standard quantization procedure~\cite{Vool2016}, we introduce a set of generalized flux variables 
\begin{equation} 
\Phi_\eta (t)=\int_{-\infty}^t ds \, V_\eta(s), \qquad \eta \in \{r,1,2\}, 
\end{equation}
where $ V_\eta $ is the voltage at the respective node. The classical equations of motion for the $\Phi_\eta$
can be derived from the Lagrangian $\mathcal{L}=T-V_{\rm tot}$, where 
\begin{equation} 
T = \frac{C_r \dot \Phi_r^2}{2} + \sum_{i=1}^N \frac{C_q (\Delta \dot \Phi_i)^2}{2}, 
\end{equation}
is the capacitive energy, while the total inductive energy, equivalent to potential energy, is given by
\begin{equation} \label{eq:circuit_potential}
\begin{split}
V_{\rm tot} &= \frac{( \Phi_r - \Phi_2 )^2}{2 L_r} 
\\
& +  \sum_{i=1}^N \left[ \frac{(\Delta \Phi_i)^2}{2 L_q} - E_J \cos \left( \frac{\Delta \Phi_i+\Phi_{\rm ext}} {\Phi_0}\right) \right].
\end{split}
\end{equation}
Here we have introduced the variables $\Delta \Phi_1\equiv \Phi_1$ and $\Delta \Phi_2=\Phi_2-\Phi_1$, which represent the phase jumps across each of the qubits. In Eq.~\eqref{eq:circuit_potential},  $\Phi_0 = \hbar/(2e) $ is the reduced flux quantum and $\Phi_{\rm ext}$ is the external flux through each of the qubit loops. In the following we set $\Phi_{\rm ext}/\Phi_0=\pi$, such that for  a Josephson energy $E_J > \Phi_0^2/L_q$ we obtain a double-well potential for the fluxes $\Delta \Phi_i$, similar to the potential considered in Sec. \ref{sec:Validity of the two-level approximation}. 

From the Lagrangian we obtain the conjugate node charges, $ Q_r = \partial \mathcal{L}/\partial \dot \Phi_r = C_r \dot \Phi_r$, and $ Q_i = \partial \mathcal{L}/\partial \Delta \dot \Phi_i = C_q \Delta \dot \Phi_i$, which simply correspond to the charges on the individual capacitors. By introducing the dimensionless variables $\phi_r=\Phi_r/\Phi_0 $,  $\phi_i=\Delta \Phi_i/\Phi_0 $ and  $\mathcal{Q}_\eta=Q_\eta/(2e)$ and promoting these variables to operators obeying $[\phi_\eta, \mathcal{Q}_{\eta'}]= i\delta_{\eta,\eta'}$ we obtain the circuit Hamiltonian  
\begin{equation}\label{eq:ham_circuit_phi} 
\begin{split}
H_\Phi = & 4E_{C_r} \mathcal{Q}_r^2 +  \frac{E_{L_r}}{2} \left(\phi_r - \sum_{i=1}^N \phi_i \right)^2 
\\ 
& +   \sum_{i=1}^N  \left[ 4E_{C_q} \mathcal{Q}_i^2  + E_J \cos \left( \phi_i \right) + \frac{E_{L_q}}{2} \phi_i^2\right].
\end{split}
\end{equation}
Here we have defined the inductive energies $E_{L_{r,q}}= \Phi_0^2/L_{r,q}$ and, following the usual convention,  the capacitive  energies  $E_{C_{r,q}}= e^2/(2C_{r,q})$. By expressing $\phi_r =\sqrt[4]{2 E_{C_r}/E_{L_r}} (a^\dag+a)$ and $\mathcal{Q}_r= i\sqrt[4]{E_{L_r}/(32E_{C_r})} (a^\dag -a)$ in terms of annihilation and creation operators and by identifying $\phi_i$ and $\mathcal{Q}_i$ with the coordinate and momentum of an effective particle moving in a potential $V(\phi_i)=  E_J \cos \left( \phi_i \right) +E_{L_q} \phi_i^2/2$, Hamiltonian  $H_{\Phi}$ is  identical to Hamiltonian $H_{D}$ in the dipole gauge.  Therefore, when we perform a TLA, we obtain the extended Dicke model \eqref{eq:ham_EDM}, with $\omega_c=\sqrt{8E_{C_r}E_{L_r}}/\hbar$ and a coupling
\begin{equation}\label{eq:gD_Circuit}
g_D =  \omega_c \left( \frac{E_{L_r}}{2E_{C_r}} \right)^{\frac{1}{4}}    |\langle \tilde \varphi_0|\phi|\tilde \varphi_1\rangle|.
\end{equation}
The dipole frequency $\tilde \omega_{10}$ and the eigenstates $|\tilde \varphi_n\rangle$ are obtained from the eigenstates of the modified qubit Hamiltonian $\tilde H_q=4E_{C_q} \mathcal{Q}^2  + V(\phi)+ E_{L_r}\phi^2/2$. Note that for this circuit configuration there appear no direct qubit-qubit interactions in the Lagrangian or the corresponding equations of motion, which therefore corresponds to the case $H_{\rm dd}=0$ considered in Sec.~\ref{sec:Multi-dipole cavity QED}. 

Of course, Hamiltonian \eqref{eq:ham_circuit_phi} is not unique and we can perform as well the unitary gauge transformation $H_Q = U H_{\Phi} U^{\dag}$, where $ U = e^{-i  \mathcal{Q}_r \sum_i \phi_i } $. In this new representation we obtain  
\begin{equation}\label{eq:ham_circuit_Q}  
\begin{split}
H_Q & =  4E_{C_r} \mathcal{Q}_r^2 +  \frac{E_{L_r}}{2} \phi_r^2
\\
& +  \sum_i \left[4E_{C_q} \left( \mathcal{Q}_i - \mathcal{Q}_r  \right)^2  + V(\phi_i)\right],
\end{split}
\end{equation}
and it can be readily seen that this Hamiltonian is equivalent to the minimal coupling Hamiltonian \eqref{eq:ham_coulomb_generic} in the Coulomb gauge. After performing a TLA we obtain the Dicke model $H_{\rm DM}$ with frequency $\tilde \omega_c= \sqrt{8( E_{C_r} + NE_{C_q} )E_{L_r}  }/\hbar$ and a coupling 
\begin{equation}
g_C=  \frac{8E_{C_q}}{\hbar}\sqrt{\frac{\omega_c}{\tilde{\omega}_c}}\left(\frac{E_{L_r}}{2E_{C_r}}\right)^{\frac{1}{4}}    |\langle \varphi_0|\mathcal{Q}| \varphi_1\rangle|.
\end{equation}
The transition frequency $\omega_{10}$ and the eigenstates $|\varphi_n\rangle$ are obtained from diagonalizing the bare qubit Hamiltonian $ H_q=4E_{C_q} \mathcal{Q}^2  + V(\phi)$. Thus,  we obtain a complete analogy between the fundamental models for electric dipoles coupled to a cavity field expressed in different gauges and a circuit QED system with flux qubits expressed in terms of different circuit variables.  

\begin{figure}
\centering
	\includegraphics[width=\columnwidth]{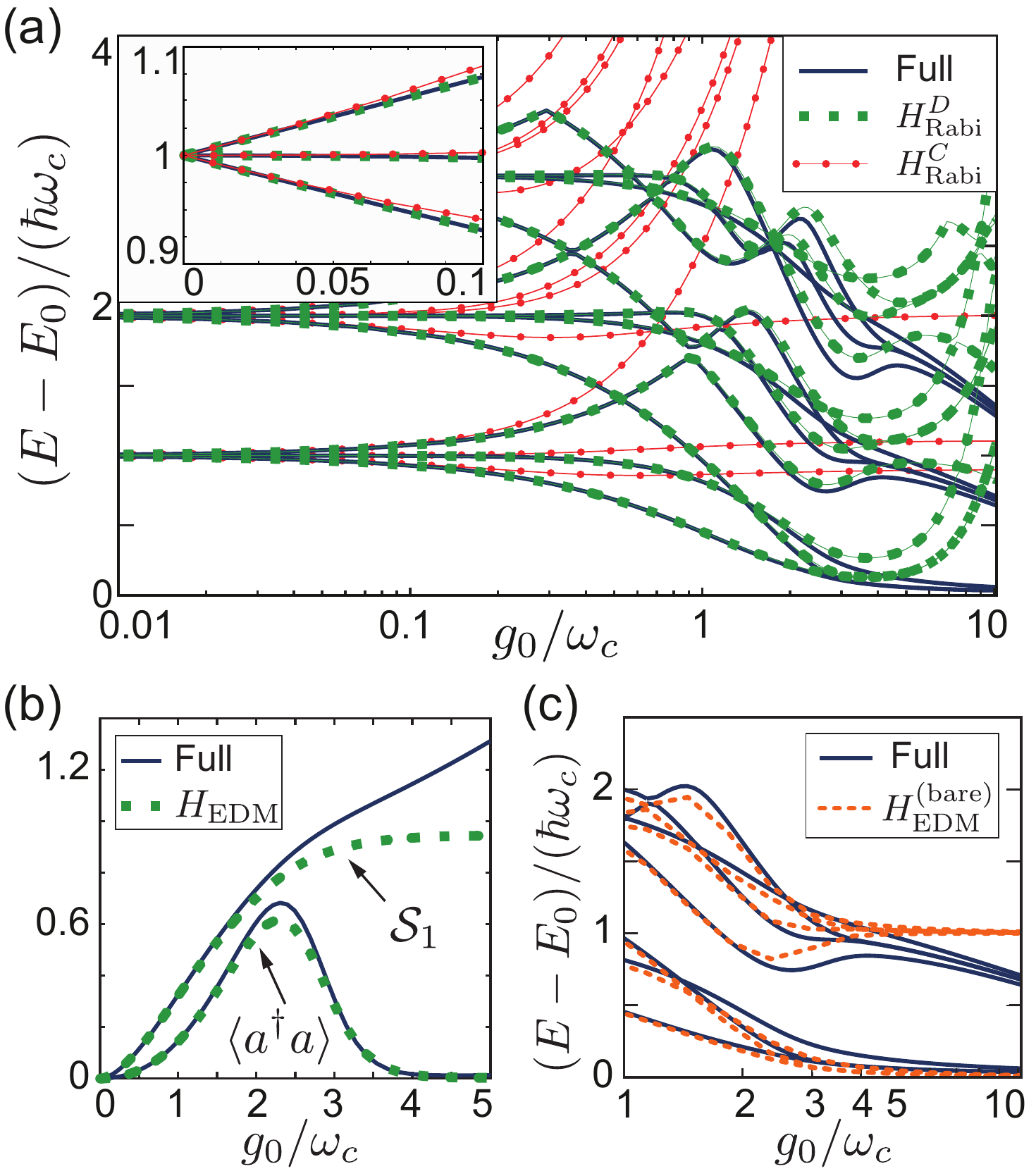}
	\caption{Two-qubit circuit QED. (a) Comparison of the energy spectra obtained from the full model $H_\Phi$ (solid blue line), the extended Dicke model $H_{\rm EDM}$ derived from $H_\Phi$ (green dashed line) and the Dicke model $H_{\rm DM}$ derived from Hamiltonian $H_{Q}$ (red dotted line) for $N=2$ flux qubits. The inset shows a zoom of the first three excitation energies for small couplings. (b) Dependence of the ground-state photon number $\langle a^\dag a\rangle$ and the single qubit entanglement entropy $\mathcal{S}_{1} = -{\rm Tr}\{\rho_1 \log_2(\rho_1)\}$ on the coupling strength $g_0$. Here $\rho_1$ is the reduced density matrix for a single qubit obtained from the density matrix of the ground state $\rho = |{\rm GS}\rangle \langle {\rm GS} |$ evaluated for the full model $H_{\Phi}$ and  for the corresponding effective model $H_{\rm EDM}$.
(c) The lowest eigenenergies (dashed orange lines) of the extended Dicke model without the $x^2$-correction,  $H^{\rm (bare)}_{\rm EDM}$,  are compared with the corresponding  energies of the full model (solid lines).
For all the plots the value of $L_r$ has been used as a tuning parameter and $C_{r}$ has been adjusted to keep the resonance condition $\omega_{10} = \omega_c=\sqrt{8E_{C_r}E_{L_r}}/\hbar$ fixed. The parameters for the flux qubits are the same as in Fig.~\ref{fig:circuit}(c).}
	\label{fig:fluxonium_spectrum_entanglement}
\end{figure}

\subsection{Few-qubit circuit QED}

At first sight it might seem more favorable to use Hamiltonian $H_Q$ as a starting point for a further simplification of this circuit. The qubit energies and eigenstates are the same as for the bare qubit and can be calculated independently of the coupling. The correction term $\sim 4E_{C_q} \mathcal{Q}_{r}^2$ can be easily absorbed into a modified resonator capacitance and the variable $\phi_r=L_r I_r$ is now directly related to the current $I_r$ through the inductor, which is a physically measurable quantity. However,  from our analysis from above we expect that due to the ``momentum"-type coupling, $H_Q$ might not permit us to make a TLA, which, in contrast, should be possible for $H_{\Phi}$. 

To confirm this intuition, we plot in Fig.~\ref{fig:fluxonium_spectrum_entanglement}(a)  the predicted energy levels $E_n$ obtained from the reduced models $H_{\rm DM}$ and $H_{\rm EDM}$ together with the exact results for the case of $N=2$ qubits. For this plot, the values for $E_J$, $E_{L_q}$ and $E_{C_q}$ have been chosen such that the frequency $\omega_{10}\approx 3$ GHz and  nonlinearity parameter $\Delta_{\rm nl} \approx 15$ are consistent with actual experimental values~\cite{Chiorescu2003}. The spectrum is plotted as a function of $g_0\sim \sqrt[4]{1/{L_r}}$, which corresponds to the coupling given in Eq.~\eqref{eq:gD_Circuit}, but evaluated for the bare qubit states $|\varphi_n\rangle$. To obtain a direct comparison with the previous results, we use $L_r$ as a tuning parameter for the coupling, but also adjust the capacitance $C_r$ to keep the resonance condition $\omega_c=\omega_{10}$ fixed. We see again very clearly the invalidity of the TLA for the charge-coupled Hamiltonian $H_{Q}$, while a good agreement between $H_{\rm EDM}$ and the full model is found.  Note that compared to the example presented in Fig.~\ref{fig:dw_combined}, the nonlinearity is now considerably smaller and therefore the discrepancy between the full Hamiltonian and $H_{\rm EDM}$ becomes visible already at $g_0/\omega_c \gtrsim 3$. Nevertheless, up to these values the effective two-level model still reproduces very well the expected separation of the spectrum into $2^N$-fold degenerate manifolds~\cite{Jaako2016}, which is not at all captured  by the spectrum of the Dicke model. Fig.  \ref{fig:fluxonium_spectrum_entanglement}(b) shows that even beyond this regime,  characteristic USC effects, such as the formation of entangled subradiant ground states and the decoupling of the cavity mode~\cite{Jaako2016,Debernardis2018}, are accurately captured by the reduced cavity QED Hamiltonian, if derived in the correct gauge.

\subsection{The $x^2$-correction}\label{subsec:X2correction}
From Fig.~\ref{fig:fluxonium_spectrum_entanglement}(a) we see that even for  very non-linear flux qubits, the spectrum of $H_{\rm EDM}$ starts to deviate significantly from the exact energies already for $g_0/\omega_c\approx 3$.  A closer inspection shows that this deviation arises mainly from the ``$x^2$-correction", i.e., the additional term $\sim E_{L_r}\phi^2/2$ in the effective qubit potential $\tilde V(\phi)$. For large couplings, this term induces a substantial modification of the qubit potential and thereby affects the coupling $g_D$ and even more strongly the qubit frequency $\tilde \omega_{10}$. While the full inclusion of this strong modification into the qubit Hamiltonian might seem to be the most accurate approach to derive a reduced two-level Hamiltonian,  Fig.~\ref{fig:fluxonium_spectrum_entanglement}(c) illustrates that this is in general not the case. In this plot  we have evaluated the spectrum of the extended Dicke model $H_{\rm EDM}^{\rm (bare)}$, which is derived from the unperturbed states $|\varphi_n\rangle$ and eigenfrequencies $\omega_n$ of the bare potential $V(\phi)$, i.e., omitting the $x^2$-correction completely.  We see that the upward-bending of the energy levels disappears and that apart from a gradual decrease of the photon frequency in the full model, $H_{\rm EDM}^{\rm (bare)}$ reproduces the qualitative features of the spectrum much more  accurately. 

To understand this somewhat counterintuitive  result, one has to keep in mind that the full interaction between the qubits and the resonator is in total given by the sum of the following three terms
\begin{equation}
H_{\rm int}=  -E_{L_r} \phi_r \sum_i \phi_i   +   \frac{E_{L_r}}{2}\sum_{i\neq j} \phi_i\phi_j +   \frac{E_{L_r}}{2}\sum_{i} \phi_i^2.
\end{equation}
By including only the last term,  i.e., the local $x^2$-correction exactly, but projecting the first two contributions onto the two-level subspace, one treats these three contributions on an unequal footing. This asymmetry can introduce unphysical artifacts in the resulting effective Hamiltonians, once the coupling to energetically higher energy levels becomes relevant. 

From our numerical studies we find that also for other nonlinear potentials, the omission of the $x^2$-correction in the derivation of the Rabi- and the extended Dicke model leads to much better qualitative predictions in the regime $g_0/\omega_c>1$. We emphasize that this is not a general result and must be verified case by case. For example, for  harmonic-like potentials the inclusion of the $x^2$ term is essential and for finite-range molecular potentials the omission of this term can even lead to unbounded ground-state energies~\cite{Rokaj2018}. Nevertheless, the results in Fig.~\ref{fig:fluxonium_spectrum_entanglement}(c) show that in particular in circuit QED, effective two-level models can be more accurate than expected from standard derivations. Importantly, even in regimes where the TLA does no longer provide accurate quantitative predictions, the discrepancies arise mainly from the effective parameters $g_D,\tilde \omega_{10}$ and $\omega_c$ that enter the extendend Dicke model, but not so much from the structure of the model itself. In particular, the observed ordering of the exact and approximate energy levels in Fig.~\ref{fig:fluxonium_spectrum_entanglement}(c) is still the same and very different from the ladder of two-fold degenerate energy levels predicted by the Dicke model under the same conditions. 

\subsection{Discussion: Circuit QED}

In circuit QED, USC conditions have been demonstrated with single flux qubits that are coupled inductively to single- or multi-mode microwave resonators~\cite{Niemczyk2010,Forndiaz2010,Baust2016,Forndiaz2017,Yoshihara2017,Chen2017}. For the quantization of such circuits one usually follows the standard approach outlined above, which results in circuit Hamiltonians similar to $H_{\Phi}$ given in Eq.~\eqref{eq:ham_circuit_phi}.  Therefore, for flux-coupled circuits one naturally obtains  a ``position-type" interaction $\sim \phi$, which permits a TLA for a sufficiently anharmonic spectrum.  Note, however, that when modeling such circuits, the usual approach of including the $\phi^2$-correction from the coupling  into a renormalization of the qubit potential $\tilde V(\phi)$ may lead to erroneous results in the regime $g_0/\omega_c>1$.  

Recently, very large couplings of about $g_0/\omega_c\approx 0.4$ have also been realized with transmon qubits that are coupled capacitively to a transmission line resonator~\cite{Bosman2017}, in which case one obtains a ``momentum-type" interaction $\sim\mathcal{Q}$. Therefore, apart from various multi-mode corrections that have already been analyzed for this setup~\cite{Ripoll2015,Malekakhlagh2016,Malekakhlagh2017,Gely2017,Munoz2018}, also the TLA must be  questioned. For conventional transmon qubits, where $E_J/E_C\gg1$, the potential $V(\phi)$ is only weakly anharmonic and the oscillator strength for the lowest transition is almost saturated ($f\approx 0.99$ for $E_J/E_C=20$). Therefore, the error introduced by making a TLA should still remain small as long as only weak excitations and moderately strong couplings are considered. However, in this transmon limit the coupling parameter is bounded by $\zeta <1$~\cite{Jaako2016,Bosman2017}, which restricts the use of this qubit design for exploring USC physics. 
 
In the other limit of a Cooper pair box~\cite{Makhlin2001}, where $E_C\gg E_J$, the electrostatic energy is the dominant energy scale and states $|\mathcal{Q}=m\rangle$ with a different number of  $m=0,\pm 1,\pm 2, ...$  Cooper pairs become energetically well-separated. In this regime a two-level subspace can be isolated by biasing the superconducting island to a charge-degeneracy point where, for example, the states $|\mathcal{Q}=0\rangle$ and $|\mathcal{Q}=1\rangle$ have the same electrostatic energy. These two states are then mixed by Josephson tunneling, resulting in the qubit states $\ket{\downarrow,\uparrow}=(|\mathcal{Q}=0\rangle\pm |\mathcal{Q}=1\rangle)/\sqrt{2}$.  Therefore, although dealing with a ``momentum"-type capacitive coupling to a microwave resonator  $\sim\mathcal{Q}$, this interaction does not couple the qubit subspace to energetically higher lying charge states and  a TLA is again well justified~\cite{Jaako2016,Manucharyan2017}.  Note, that due to the discreteness of the charge states and the presence of a bias voltage, the energy levels in a Cooper pair box can no longer be directly compared with a regular particle moving in a potential well. In particular, for this system the TRK sum rule and the relation between matrix elements of $\mathcal{Q}$ and $\phi$ similar to Eq.~\eqref{eq:XPrelation} do no longer apply. In this parameter regime the strict analogy between circuit QED and cavity QED with regular dipoles fails.

\section{Conclusions}
\label{sec:conclusions}
In summary, we have discussed the crucial role of the choice of gauge in the derivation of effective models for light-matter interactions in the USC regime.  Specifically, we have shown that in the Coulomb gauge the couplings to higher excited states of the dipole potential are in general not energetically suppressed, and even for very anharmonic potentials performing the TLA can give completely wrong results. While for harmonic dipoles or potentials where the oscillator strength is almost saturated, i.e., $f\approx 1$, the Coulomb gauge and the dipole gauge still give very similar results in the collective USC regime, significant deviations are found for potentials with $f<1$ and, more generally, in the single-dipole USC regime. Under such conditions not only the effective parameters, but also the structure of the effective cavity QED models depends on the chosen gauge. Thus the findings of this work have an immediate relevance for various USC cavity QED experiments, for example, intersubband polaritons in asymmetric wells or circuit QED devices.

We emphasize that in the current work we have focused on the multi-level structure of the matter system, assuming the coupling to a single electromagnetic resonance. This is justified in essentially single-mode photonic cavities, as lumped-elements resonators in the microwave domain \cite{Jaako2016}. In the case of generic resonators though, other photonic modes are present, and neglecting them can lead to unphysical predictions, like superluminal signal propagation~\cite{Munoz2018}, when the light-matter coupling becomes non-perturbative.

\emph{Note added.}---After submission of this work a related study about the TLA in different gauges appeared~\cite{Stokes2018}.  

\acknowledgements
This work was supported by the Austrian Science Fund (FWF) through the SFB FoQuS, Grant No. F40, the DK CoQuS, Grant No. W 1210, and the START Grant No.  Y 591-N16.
SDL acknowledges support from a Royal Society research fellowship.

\appendix

\section{Counter-no-go theorem}
\label{app:Generalised counter no-go theorem}
In this Appendix we provide a more general derivation of the counter-no-go theorem for the coupling parameter $\zeta_D$ discussed in Sec. \ref{sec:counter_no_go}. We first remark that from the TRK sum rule we obtain both a lower and an upper bound on the coupling parameter,
\begin{equation}\label{eq:zetaD_Bounds}
\frac{|\tilde{x}_{10}|^4}{x_D^4}  \leq \zeta_D \leq \frac{D^2}{\tilde{\omega}_{10}^2},
\end{equation}
where $x_D = \sqrt{ \hbar/(2mD)} $ is the harmonic oscillator length in the limit $D\sim g_0 \rightarrow \infty$. For finite, but large $D$ the lowest eigenstates of the total potential  $\tilde{V}(x) = V(x) + mD^2x^2/2$ are localized around $x\approx 0$ and we can expand the bare potential as
\begin{equation}
V(x)\simeq \frac{c_{2n}}{(2n)!} x^{2n}.
\end{equation}
Here we have assumed that the potential is symmetric and that $c_{2n}$ is the lowest non-vanishing coefficient in the Taylor series of $V(x)$. By considering $V(x)$ as a small correction to $mD^2 x^2/2$, we can use perturbation theory and obtain
\begin{equation}
\tilde{\omega}_{10} \simeq   D + \frac{c_{2n}}{\hbar}\frac{\mathcal{N}_0}{2n!} x_D^{2n},
\end{equation}
and 
\begin{equation}\label{eq:x01_USC_perturb_dipolarGauge}
\tilde{x}_{10}\simeq x_D - \left(\frac{ c_{2n}}{2\hbar D}\frac{\mathcal{N}_0}{2n!}  \right) x_D^{2n+1}.
\end{equation}
where $\mathcal{N}_0 = \langle{1 | (a + a^{\dag})^{2n}) | 1 \rangle} - \langle{0 | (a + a^{\dag})^{2n}) | 0 \rangle}$ is a positive normalization constant.
We see that for any symmetric potential that is anti-confining at the origin, i.e., $c_{2n}<0$, there is a certain value of the coupling $g_0$ beyond which $\tilde x_{10}>x_D$ and therefore $\zeta_D>1$. For a confining potential, where $c_{2n}>0$, we find $\zeta_D<1$ in the large coupling limit, but there might still be intermediate coupling regimes where the value of $\zeta_D$ exceeds the value of one.

\section{Exact diagonalization of $H_C$ in the limit $N\rightarrow \infty$}
\label{app:no-go_dep_shift}
In the dilute regime, where the number of dipoles $N$ is much larger than the average number of excitations, we can use a multi-level Holstein-Primakoff approximation to calculate the excitation energies of the full Hamiltonian $H_C$ given in Eq.~\eqref{eq:ham_coulomb_generic}. Under this approximation we obtain~\cite{ViehmannPRL2011,Hopfield1958} 
\begin{equation}
\begin{split}
H_{C} & \simeq  \tilde \omega_c a^{\dag}a + \sum_n \omega_n b^{\dag}_n b_n \\
&- \frac{G_C}{2}\left(a+a^{\dag}\right)\sum_n \nu_n \left( b_n + b_n^{\dag} \right)
\end{split}
\end{equation}
where $\tilde \omega_c=\sqrt{\omega_c^2+ND^2}$, $D^2 = \hbar g_0^2/( 2m x_{10}^2 \omega_c )$ and $G_C=\sqrt{N} g_C = G_0 \omega_{10}/\sqrt{\omega_c \tilde{\omega}_c}$. Here the operators $b^\dag_n=1/\sqrt{N}\sum_{i=1}^N |n_i\rangle \langle 0_i |$ create a collective excitation in the $n$-th energy level of the bare dipole Hamiltonian and $\nu_n = (x_{n0}/x_{10})(\omega_{n0}/\omega_{10})$. In the low excitation limit we can neglect double occupancies of states other than the ground state and treat the $b_n$ as bosonic operators with commutation relations $[b_n, b_m^{\dag}] \simeq \delta_{nm}$.
The eigenfrequencies $\omega$ of this system are then given by the solutions of the equation
\begin{equation}\label{eq:polariton_low_energy_spec_full}
\omega^2 + G_0^2\sum_n \frac{\nu_n^2\omega_{10}^2/(\omega_{n0} \omega_c)}{1-\omega^2/\omega_{n0}^2} = \omega_c^2 + ND^2.
\end{equation}
This equation can be solved numerically and the lowest two eigenfrequencies denoted by $\omega_{\pm}$ are plotted as solid lines in Fig.~\ref{fig:HP_polariton_modes}(a) and (b). 
Since all levels of the dipole potential are included, the spectrum obtained from this equation is gauge invariant, which can be verified by repeating the same calculation in the electric dipole gauge, starting from Hamiltonian~\eqref{eq:ham_many_dip_dipgauge}.

By looking only at the lowest solution of Eq.~\eqref{eq:polariton_low_energy_spec_full}, we can assume that $\omega_- \ll \omega_{n0}$ and obtain the approximate result
\begin{equation}
\omega_{-}^2 \simeq \frac{ \omega_c^2+ ND^2\left(1  - \frac{2m}{\hbar}\sum_n x_{n0}^2\omega_{n0} \right) }{ 1 + G_0^2\sum_n \frac{\nu_n^2\omega_{10}^2}{\omega_{n0}^3\omega_c}}.
\end{equation}
From the TRK sum rule it follows that the term in the parentheses is zero and 
\begin{equation}
\lim_{G_0 \rightarrow \infty} \omega_{-} = 0.
\end{equation}
Therefore, the lower polariton frequency approaches zero  for large enough coupling. This finding contradicts the finite value of $\omega_-$ in Eq.~\eqref{eq:DepolarizationShift}, as obtained from the Dicke model in the Coulomb gauge.

\end{document}